\documentclass[fleqn,usenatbib]{mnras}

\usepackage{newtxtext,newtxmath}

\usepackage[T1]{fontenc}

\DeclareRobustCommand{\VAN}[3]{#2}
\let\VANthebibliography\thebibliography
\def\thebibliography{\DeclareRobustCommand{\VAN}[3]{##3}\VANthebibliography}

\usepackage{graphicx}	
\usepackage{lscape}
\usepackage{longtable}

\newcommand{\myr}{$\rm mas\,yr^{-1}$}

\title[The vertex coordinates]{The vertex coordinates of the Galaxy's stellar systems according to the Gaia DR3 catalogue}

\author[Dmytrenko et al.]{
A. M. Dmytrenko,$^{1}$\thanks{E-mail: astronom.karazin007@gmail.com (AMD)}
P. N. Fedorov,$^{1}$\thanks{E-mail: pnfedorov@gmail.com (PNF)}
V. S. Akhmetov,$^{1,2}$\thanks{E-mail: akhmetovvs@gmail.com (VSA)}
A. B. Velichko$^{1,3}$
and S. I. Denyshchenko$^{1}$
\\
$^{1}$Institute of astronomy of V.N.Karazin Kharkiv national university, Svobody sq. 4, 61022, Kharkiv, Ukraine\\
$^{2}$INAF-Osservatorio Astrofisico di Torino, Via Osservatorio 20, Pino Torinese, Turin, I-10025, Italy\\
$^{3}$Department of Astronomy, University of Geneva, Chemin Pegasi 51, 1290 Versoix, Switzerland
}

\date{Accepted XXX. Received YYY; in original form ZZZ}

\pubyear{2023}

\begin{document}
\label{firstpage}
\pagerange{\pageref{firstpage}--\pageref{lastpage}}
\maketitle

\begin{abstract}
We present the results of determining the coordinates of the vertices of various stellar systems, the centroids of which are located in the Galactic plane. To do this, the positions, parallaxes, proper motions, and radial velocities of red giants and subgiants contained in the $Gaia$~DR3 catalogue have been used. When determining the components of the deformation velocity tensors in local coordinate systems, we found the coordinates of the vertices of the stellar systems under study. It turned out that there is a complex dependence of vertex deviations $l_{xy}$ in Galactocentric cylindrical ($R$, $\theta$) and Galactic rectangular ($X,Y$) coordinates.
Based on the approach proposed in this paper, heliocentric distances to vertices have been determined for the first time. The results obtained show that in addition to the fact that the angular coordinates of the Galactic center and the vertices of stellar systems do not coincide, their heliocentric distances do not coincide as well. This presumably indicates that there are structures in the Galaxy that noticeably affect its axisymmetry.
\end{abstract}


\section{Introduction}
\label{sec:intro}

The third release of the $Gaia$ mission, $Gaia$~DR3 data (\cite{Prusti2016, Vallenari2022}), made new data available for studying the stellar kinematics not only in the Solar neighborhood, but also in a significant part of the Milky Way. The availability of high-precision data of millions of stars in the releases of the $Gaia$ mission makes it possible to obtain new information about the kinematics of stars.

Particularly valuable for kinematic studies is the availability of data on radial velocities and parallaxes, which, together with the stellar proper motions, make it possible to analyze the three-dimensional velocity field ${\bf V}({\bf r})$. This entails the emergence of new possibilities for determining some global kinematic parameters, as shown in the works by \cite{Fedorov2021, Fedorov2023}. In this work, we present the results of determining the kinematic centers of rotation of various stellar systems. And although there are certain difficulties in interpreting the results obtained, caused, for example, by the discrepancy between the distances to sources determined from the $Gaia$ parallaxes and using the Bayesian method, usage of different estimates of the distances $R_\odot$ to the  Galactic center, or relatively small number of stars with known radial velocities ($\sim$33.7 million), the relevance of such works is beyond doubt.

In this paper, we determine the vertex coordinates of stellar samples whose centroids are located in the Galactic plane, using 3 different parallax sets. These are trigonometric parallaxes given in $Gaia$ DR3, the same parallaxes, but corrected using the Bayesian method (\cite{Bailer-Jones2021}), as well as parallaxes corrected with the use of the Parallax bias $Z_5$ proposed by \cite{Lindegren2021}. In addition, according to the recommendations from \cite{Cantat-Gaudin2021}, the proper motions of our sample were corrected in the range of magnitudes $M_G$ 9 -- 13. One way to determine the coordinates of the vertex is to analyze the strain rate tensor of the stellar system (\cite{Bobylev2004, Bobylev2020}).

The paper is structured as follows. In section \ref{sec:sample}, we describe the basic steps to form stellar samples in a rectangular coordinate system from giants and subgiants, which are contained in the $Gaia$~DR3 catalogue.
In section \ref{sec:vertexcoords} we present the formulas that are used in this paper to calculate the angle $l_{xy}$.
Section \ref{sec:analysis} contains a description of the problem solution, analysis and interpretation of the results obtained.
\section{The sample}
\label{sec:sample}

It is well known, that the Galactic rectangular coordinate system with the origin at the barycenter of the Solar System is defined by a right-handed triple of mutually orthogonal unit vectors $( \bf i, \bf j, \bf k)$ directed as follows: the X axis from the observer towards the galactic center $L=0^\circ$, $B=0^\circ$, the axis Y in the direction of Galactic rotation $L=90^\circ$, $B=0^\circ$, the Z axis is parallel to the direction to the North Pole of the Galaxy $B=90^\circ$. As noted in (\cite{Fedorov2021, Fedorov2023}), a similar coordinate system can be introduced at any arbitrary point on the Galactic plane, provided that the spatial coordinates and components of the spatial velocity are known for each star. The transition from the Galactic Cartesian coordinate system with the origin at the barycenter of the Solar System $(XYZ)$ to such a local Cartesian system $(xyz)$ with the origin at the chosen point $(xyz)$ is equivalent to moving a fictitious observer from the barycenter of the Solar System to the point specified by the coordinates of the chosen origin of the system. In the local Cartesian coordinate system, as well as in the Galactic Cartesian system, the x-axis ($l=L=0^\circ$, $b=B=0^\circ$) is always directed from a particular centroid to the center of the Galaxy, the $Oy$-axis ($l=L=90^\circ$ , $b=B=0^\circ$) in the direction of galactic rotation and perpendicular to $Ox$, while the $Oz$-axis ($b=B=90^\circ$) is always perpendicular to the plane of the Galaxy. The orientation of the x and y axes of the local coordinate systems was specified using the value R$_\odot$ = 8.28 kpc. (\cite{Abuter2021}). In a particular case, for an observer who is in the Sun, the local coordinate system will coincide with the rectangular Galactic coordinate system.

In this work, 33 million stars from $Gaia$~DR3 were selected for which the radial velocities are known. From this sample, as in our previous work (\cite{Fedorov2023}), were excluded those stars for which the following conditions are satisfied (\cite{Lindegren2018}):
\begin{equation*}
 \begin{cases}
   RUWE > 1.4,\\
   \pi/\sigma_\pi < 5,\\
   (\mu_\alpha/\sigma_{\mu_\alpha})^2 + (\mu_\delta/\sigma_{\mu_\delta})^2 < 25.
 \end{cases}
\end{equation*}

\begin{figure}
   \centering
\resizebox{\hsize}{!}
   {\includegraphics{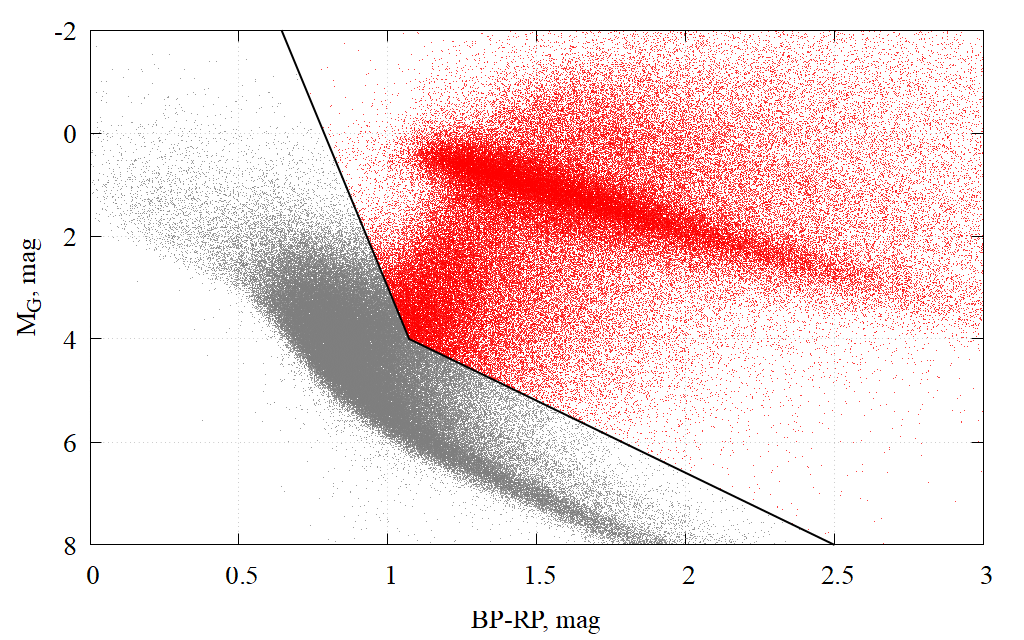}}  
   \caption{Selection of red giants and subgiants.}
\label{fig:BP-RP_MG}
\end{figure}

By cutting off the main sequence on the $M_G-(BP-RP)$ Hertzsprung-Russell diagram with two linear functions, as shown in Fig.~\ref{fig:BP-RP_MG}, from the approximately 30 million stars (remaining after applying the Lindegren criteria), giants and subgiants were selected. Finally, a sample of approximately 15 million giants and subgiants was used further in kinematic studies.
The points from which, fictitious observations were made by a fictitious observer, were given as follows. Firstly, we single out spherical regions with a radius of 1~kpc, whose centers are located at the nodes of a rectangular grid coinciding with the Galactic plane. Coincidence with the Galactic plane is provided by setting the condition $Z = 0$ for the coordinates of any node. Thus, the position of each node is uniquely specified by a pair of coordinates $X$ and $Y$. The distance between adjacent nodes along both coordinates was set equal to 250~pc, and the boundaries of the entire region under study in the Galactic plane were limited by the range of heliocentric distances from -8 to +8 kpc along both coordinates. Each sphere circumscribed around a given node includes stars located at distances not exceeding 1~kpc from it. Thus, the nodes of the rectangular grid are centroids whose velocities are equal to the average velocity of the stars located within the corresponding spheres. Only spheres containing at least 500 stars have been used in the work.

\section{Determining the coordinates of the vertex.}
\label{sec:vertexcoords}

In this paper, deformation velocity tensors were computed for all stellar systems whose centers are located at the nodes of a rectangular grid, as indicated above (see also \cite{Fedorov2023}). If in the local Cartesian coordinate system the radius-vector of stars is denoted as ${\bf r} =(x,y,z)=q_i$, and their velocities are ${\bf V} =(V_x,V_y,V_z)=V_i$, then from the expansion of the velocity field $\bf{V(\bf{r})}$ in the vicinity of the node (centroid), we get:
\begin{equation}
\label{eq:sol_m}
V_i(d{\bf r}) =  V_i(0) + \frac{1}{2}\left( \frac{\partial V_i}{\partial q_k} - \frac{\partial V_k}{\partial q_i}\right)_0dq_k + \frac{1}{2}\left( \frac{\partial V_i}{\partial q_k} + \frac{\partial V_k}{\partial q_i}\right)_0dq_k,
\end{equation}
where
\begin{equation*}
\frac{1}{2}\left(\frac{\partial V_i}{\partial q_k} - \frac{\partial V_k}{\partial q_i}\right)_0 = \omega_{ik} = -\omega_{ki} = M^{-}
\end{equation*}
are components of antisymmetric and
\begin{equation*}
\frac{1}{2}\left(\frac{\partial V_i}{\partial q_k} + \frac{\partial V_k}{\partial q_i}\right)_0 = m^{+}_{ik} = m^{+}_{ki} = M^{+}
\end{equation*}
symmetric tensors respectively. $i, k$ = 1, 2, 3.

By making the appropriate substitutions, this equation can also be written in Galactic spherical coordinates. In this form, it is known in the literature as the Ogorodnikov--Milne (O--M) kinematic model (\cite{Ogorodnikov1932, Ogorodnikov1965}).
The second rank symmetric tensor $M^{+}$ is called the  deformation velocity tensor. The matrix of this tensor has the form:
\begin{equation}
{M^{+} =
\begin{pmatrix}
m^{+}_{11} & m^{+}_{12} & m^{+}_{13} \\
m^{+}_{21} & m^{+}_{22} & m^{+}_{23} \\
m^{+}_{31} & m^{+}_{32} & m^{+}_{33}
\end{pmatrix}}.
\label{eq:mptensor_3D}
\end{equation}

The $M^+$ tensor completely determines the rate of deformation motion in the stellar system under consideration. In the rectangular Galactic coordinate system, it has 9 components, 6 of which are independent. As is known, the kinematic interpretation of the diagonal components of the tensor $m^+_{11}$, $m^+_{22}$, $m^+_{33}$ is that these quantities are the velocities of relative elongation (contraction/expansion) along the axes of the coordinate system, and the non-diagonal components $m^+_{12} = m^+_{21}$, $m^+_{13}, = m^+_{31}$, $m^+_{23} = m^+_{32}$, characterize the velocities of angular deformation in the planes $(xOy)$, $(yOz)$ and $(xOz)$, respectively. The velocity of angular deformation is understood as a change in the right angle in these planes as a result of deformation.

It is convenient to present the results of calculations in the cylindrical Galactocentric coordinate system $R,\theta,Z$, since its unit vectors are parallel to the unit vectors of local rectangular coordinate systems in each centroid. For example, a node centered on the Sun will have the following coordinates: $R=R_\odot=8.28$ kpc, $\theta=180^\circ$, $Z=0$ kpc.

\begin{figure}
   \centering
\resizebox{\hsize}{!}
   {\includegraphics{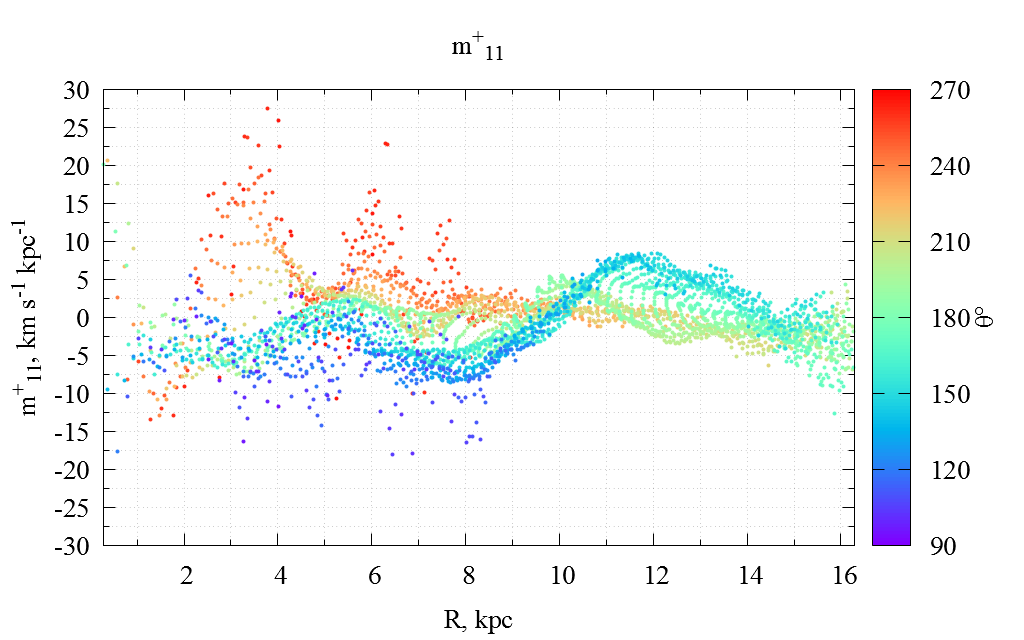}}
\resizebox{\hsize}{!}
   {\includegraphics{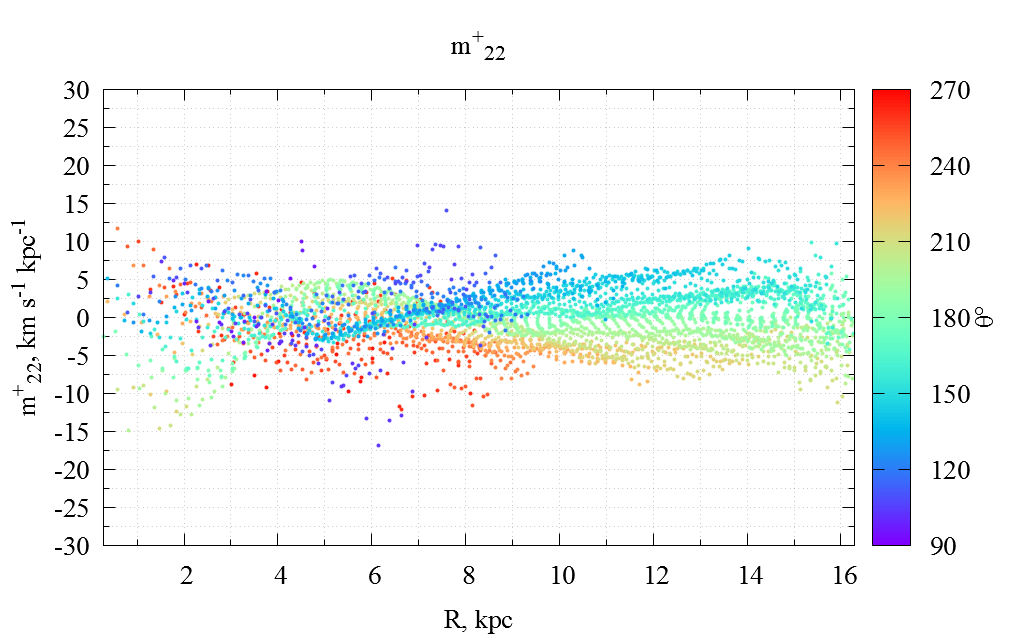}} 
\resizebox{\hsize}{!}
   {\includegraphics{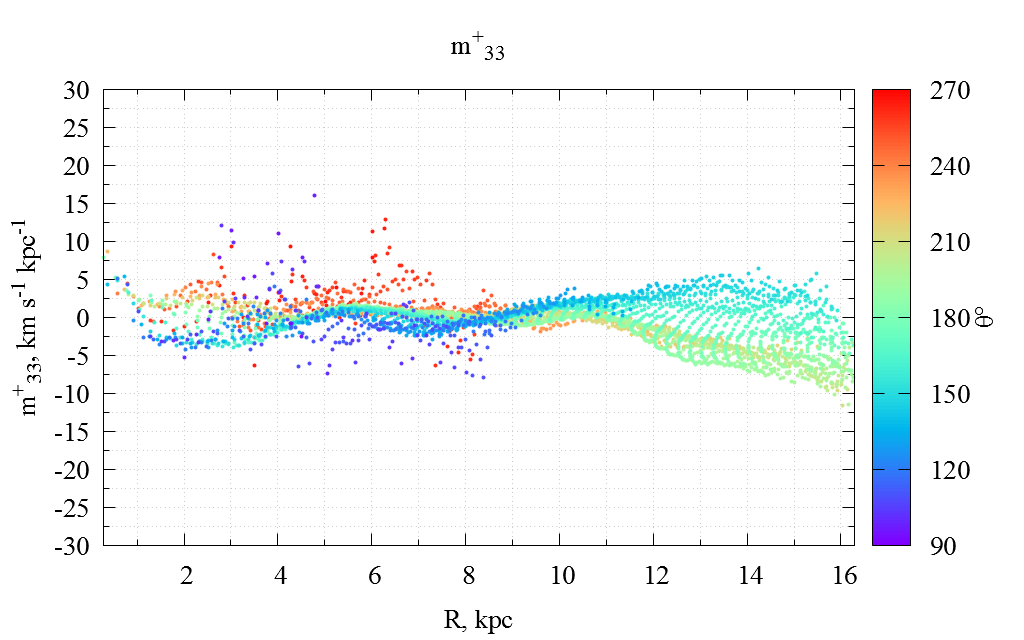}}       
   \caption{Diagonal components of the $M^+$ tensor as a function of Galactocentric cylindrical coordinates.}
\label{fig:Mp_diag}
\end{figure}

\begin{figure}
   \centering
\resizebox{\hsize}{!}
   {\includegraphics{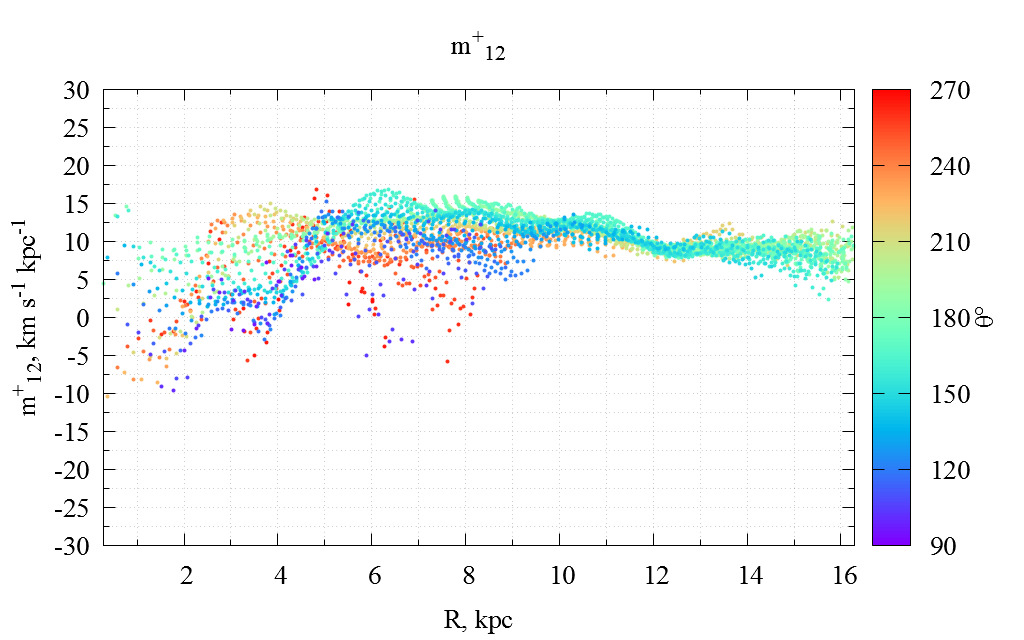}}
\resizebox{\hsize}{!}
   {\includegraphics{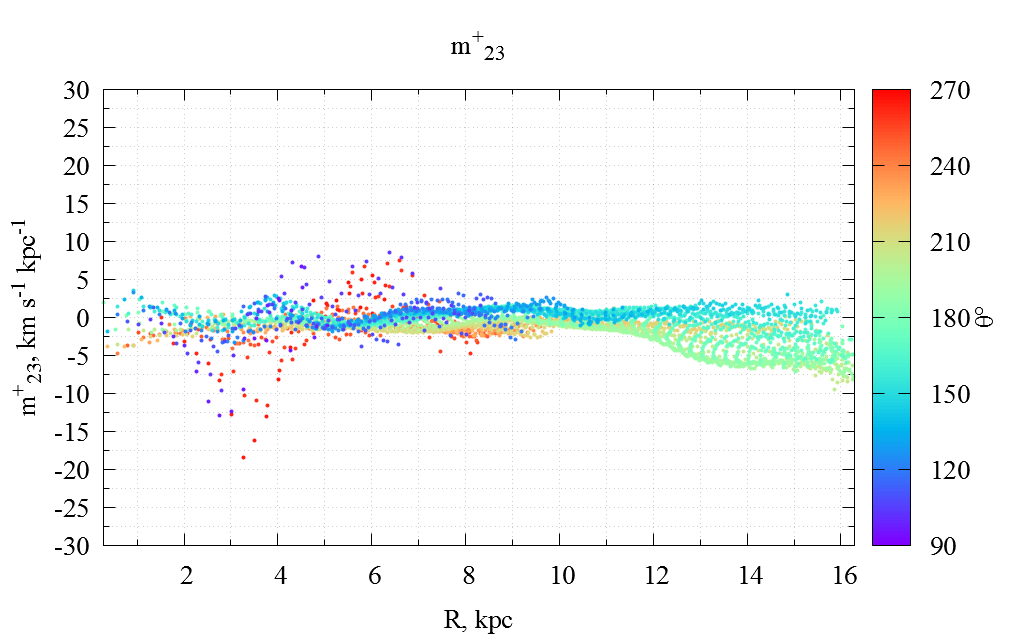}} 
\resizebox{\hsize}{!}
   {\includegraphics{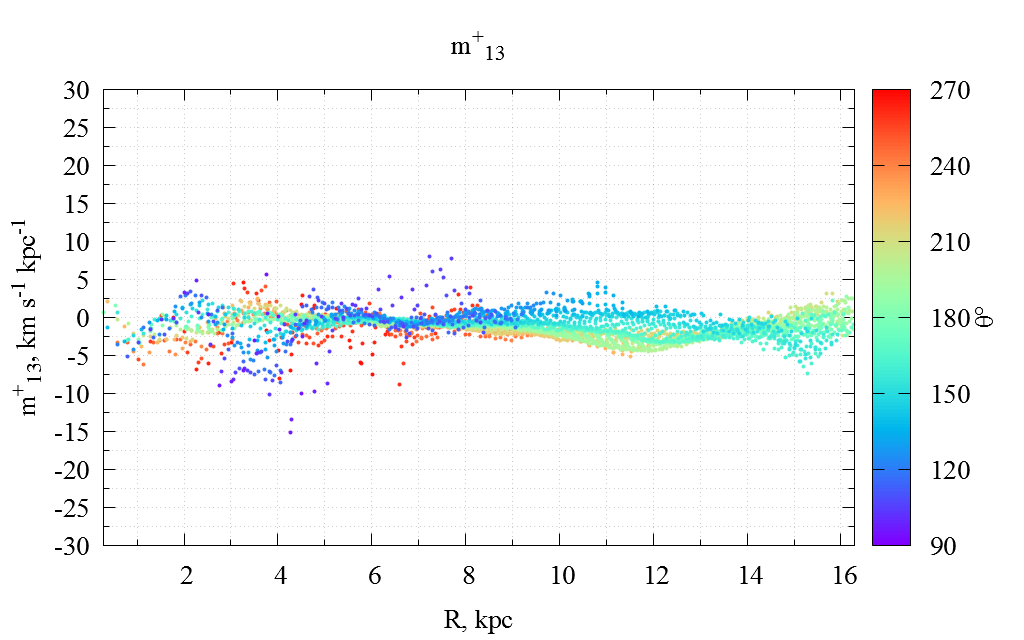}}       
   \caption{Non-diagonal components of the $M^+$ tensor as a function of Galactocentric cylindrical coordinates.}
\label{fig:Mp_nondiag}
\end{figure}

Fig.~\ref{fig:Mp_diag} shows the dependencies of the diagonal components of the deformation velocity tensors $m^+_{11}$, $m^+_{22}$, $m^+_{33}$ in the Galactocentric cylindrical coordinates $R$. These dependencies are additionally color-coded for different values of the coordinate $\theta$.
Fig.~\ref{fig:Mp_nondiag} shows the dependencies of the non-diagonal components of the velocity deformation tensors $m^+_{12}$, $m^+_{13}$, $m^+_{23}$ on $R$ and $\theta$.
It is known from continuum mechanics (\cite{Tarapov2002}) that no matter how a particle of a continuous medium moves, all its deformation can be reduced to the simplest - expansion (contraction) along three mutually perpendicular directions, which are the main axes $x'y'z'$ of the tensor. In the system of its main axes $x'y'z'$, the velocity deformation tensor $M^+$ will have the following matrix:
\begin{equation}
{M^{+} =
\begin{pmatrix}
\lambda_1 & 0 & 0 \\
0 & \lambda_2 & 0 \\
0 & 0 & \lambda_3
\end{pmatrix}}.
\label{eq:mainaxiestensor_3D}
\end{equation}
where the diagonal components $\lambda_1$, $\lambda_2$, $\lambda_3$ are the principal values of the tensor, which are determined by solving the characteristic cubic equation:
\begin{equation}
{M^{+} =
\begin{pmatrix}
m^{+}_{11}-\lambda_1 & m^{+}_{12} & m^{+}_{13} \\
m^{+}_{21} & m^{+}_{22}-\lambda_2 & m^{+}_{23} \\
m^{+}_{31} & m^{+}_{32} & m^{+}_{33}-\lambda_3
\end{pmatrix}}.
\label{eq:mptensorwithlambdas_3D}
\end{equation}
If $\lambda_1$, $\lambda_2$, $\lambda_3$ are positive, then the tensor surface is an ellipsoid, if $\lambda_1$, $\lambda_2$, $\lambda_3$ have different signs, the tensor surface is a hyperboloid.

\begin{figure}
   \centering
\resizebox{\hsize}{!}
   {\includegraphics{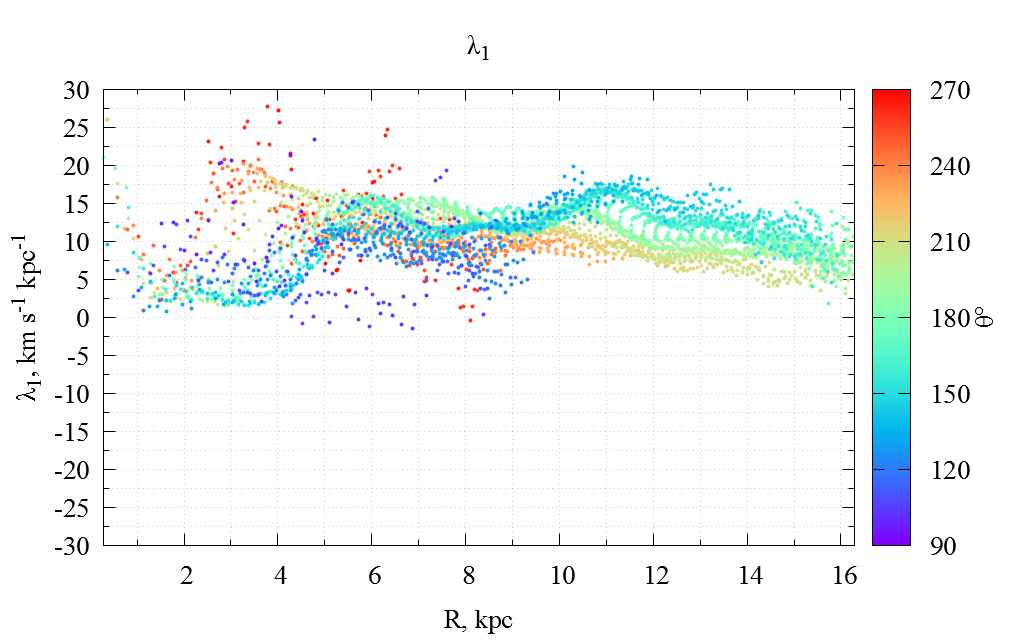}}
\resizebox{\hsize}{!}
   {\includegraphics{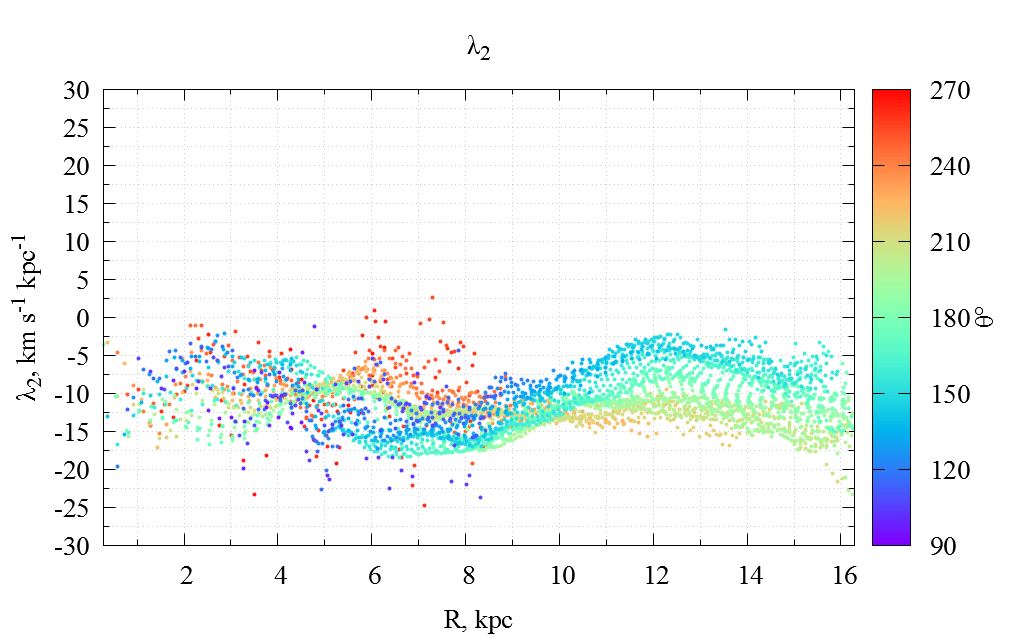}} 
\resizebox{\hsize}{!}
   {\includegraphics{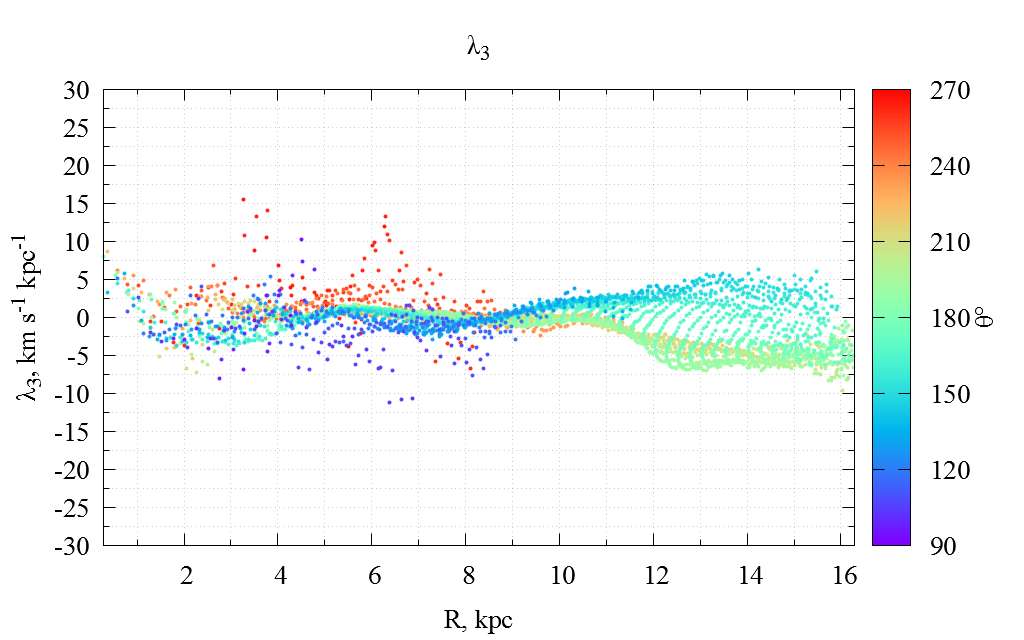}}       
   \caption{Eigenvalues of the $M^+$ tensor as a function of Galactocentric cylindrical coordinates.}
\label{fig:Mp_lambdas}
\end{figure}

Fig.~\ref{fig:Mp_lambdas} shows the dependencies of the principal values of the tensor in the Galactocentric cylindrical coordinates $R$ and $\theta$.
As can be seen from the figures, $\lambda_1$ and $\lambda_2$ have opposite signs and different behavior, while the value of $\lambda_3$ is almost independent of $R$ and $\theta$ and is close to zero on average. Only at the Galactocentric distance greater than 11 kpc do we see a slight systematic deviation of $\lambda_3$ from zero, which we neglected in this work. The numerical values of the parameters $\lambda_1$, $\lambda_2$, $\lambda_3$ indicate that the velocity deformation tensor in the principal axes is almost independent of the $Z$ coordinate and, therefore, is very close to a flat (two-dimensional) tensor.

The kinematic interpretation of the components $\lambda_1$, $\lambda_2$, $\lambda_3$ in the system of principal axes $x'y'z'$ is identical to that of $m^+_{11}$, $m^+_{22}$, $m^+_{33}$ in the system of axes $xyz$. These quantities are the velocities of relative elongations (contractions/expansions) along the principal axes of the tensor. As can be seen from the figures, during the transition from the Galactic coordinate system to the system of principal axes, the $m^+_{33}$ component almost exactly transforms into $\lambda_3$, i.e., these two components practically coincide numerically. Indeed, it turned out that $m^+_{33}$ almost coincides with $\lambda_3$. However, it is clearly seen that the components $m^+_{11}$ and $m^+_{22}$ were not completely transformed into $\lambda_2$ and $\lambda_1$. This is due to the fact that the $m^+_{12}$ component, being non-zero, contributes to $\lambda_2$ and $\lambda_1$. Considering the components $\lambda_2$ and $\lambda_1$ in the range of 4-12~kpc, one can see that the dependence of $m^+_{11}$ on $R$ and $\theta$, similar to a sinusoidal one, has been transformed into an almost analogous dependence of the component $\lambda_2$. While the dependence $m^+_{22}$ on $R$ and $\theta$ has been transformed into a similar dependence $\lambda_1$. These facts indicate that the main axes $Ox'$ and $Oy'$ of the  deformation velocity tensors are located almost in the Galactic plane, and the $Oz'$ axis practically coincides with the $Oz$ axis. This is also confirmed by the behavior of the $z$-dependent parameters $m^+_{33}$, $m^+_{13}$ and $m^+_{23}$, which were presented above. The values of these parameters in the entire range of $R$ and $\theta$ are close to zero and indicate that the motion of stars is predominantly parallel to the Galactic plane. Therefore, in the further analysis, we consider the tensor $M^+$ as flat, i.e. having only four components $m^+_{11}$, $m^+_{12}$, $m^+_{21}$, $m^+_{22}$, three of which are independent ($m^+_{12} = m^+_{21}$):
\begin{equation}
{M^{+} =
\begin{pmatrix}
m^{+}_{11} & m^{+}_{12} \\
m^{+}_{21} & m^{+}_{22} \\
\end{pmatrix}}.
\label{eq:mptensor_2D}
\end{equation}
The components of this tensor are related to the gradients of the  velocity components  in the local rectangular Galactic ($V_x$, $V_y$) and in the Galactocentric cylindrical ($V_R$, $V_\theta$) coordinate systems along their coordinate axes by the following relations (\cite{Chandrasekhar1945, Ogorodnikov1965}):
\begin{equation}
\begin{array}{r}
A = m^{+}_{12} =
\frac{1}{2}\Big( \frac{\partial V_x}{\partial y} + \frac{\partial V_y}{\partial x}\Big) =
\frac{1}{2}\Big( \frac{1}{R} \frac{\partial V_R}{\partial \theta} - \frac{V_\theta}{R} +
\frac{\partial V_\theta}{\partial R}\Big) \\
B = \omega_{3} =
\frac{1}{2}\Big( \frac{\partial V_y}{\partial x} - \frac{\partial V_x}{\partial y}\Big) = 
\frac{1}{2}\Big( \frac{\partial V_\theta}{\partial R} - \frac{1}{R}\frac{\partial V_R}{\partial\theta} +
\frac{V_\theta}{R}\Big)  \\
C = \frac{m^{+}_{11} - m^{+}_{22}}{2} =
\frac{1}{2}\Big( \frac{\partial V_x}{\partial x} - \frac{\partial V_y}{\partial y}\Big) = 
\frac{1}{2}\Big( \frac{\partial V_R}{\partial R} - \frac{1}{R}\frac{\partial V_\theta}{\partial\theta} -
\frac{V_R}{R}\Big) \\
K = \frac{m^{+}_{11} + m^{+}_{22}}{2} =
\frac{1}{2}\Big( \frac{\partial V_x}{\partial x} + \frac{\partial V_y}{\partial y}\Big) = 
\frac{1}{2}\Big( \frac{\partial V_R}{\partial R} + \frac{1}{R}\frac{\partial V_\theta}{\partial\theta} +
\frac{V_R}{R}\Big)
\label{eq:oortparams}
\end{array}
\end{equation}

where $A$, $B$, $C$ and $K$ are parameters similar to the generalized Oort constants that can be found for each region of stars under study.

In continuum mechanics (\cite{Tarapov2002}), it is shown that for a two-dimensional tensor $M^+$, the angles $\beta_1$ and $\beta_2$, which form the main axes of the tensor $M^+$ with the axes $Ox$ and $Oy$ of the coordinate system used, are found from the expression:
\begin{equation}
tg(2\beta) = tg(2\beta_1) = tg(2\beta_2) = \frac{2 m^{+}_{12}}{m^{+}_{11} - m^{+}_{22}},
\label{eq:beta_Mp}
\end{equation}
where $\beta_1$ is the angle between $Ox$ and $Ox'$, and $\beta_2$ is between $Oy$ and $Oy'$.

If the components of the $M^+$ tensor are expressed in terms of the parameters $A$, $B$, $C$, $K$ then the matrix \ref{eq:mptensor_2D} will have the following form:
\begin{equation}
{M^{+} =
\begin{pmatrix}
K+C & A \\
A & K-C \\
\end{pmatrix}}.
\label{eq:mptensor_oorts_2D}
\end{equation}
and formula \ref{eq:beta_Mp} will be accordingly transformed to the form:
\begin{equation}
tg(2\beta) = \frac{2 A}{(K+C)-(K-C)} = \frac{A}{C}.
\label{eq:beta_oorts}
\end{equation}
In the case of a purely Oort rotation, there is only the rotational component $V_\theta$, while the components $V_z$ and $V_R$ = 0. In this case, $V_\theta$ does not depend on $\theta$, i.e.
\begin{equation}
{\frac{\partial V_R}{\partial R}=0, \:\:\:\:\:\: V_R + \frac{\partial V_\theta}{\partial\theta}=0}.
\label{eq:CKnull}
\end{equation}
As a result, the terms by which the parameters $C$ and $K$ are determined in a cylindrical Galactocentric coordinate system, are equal to zero:
\begin{equation}
{{V_R=0, \:\:\:\:\:\: \frac{\partial V_\theta}{\partial\theta}=0}},
\label{eq:CKnull_oort}
\end{equation}
and hence $C$ and $K$ are also equal to zero, and the tensor $M^+$ will have zero diagonal components:
\begin{equation}
{M^{+} =
\begin{pmatrix}
0 & A \\
A & 0 \\
\end{pmatrix}}.
\label{eq:mptensor_oorts_axisymetry_2D}
\end{equation}
Thus, for a purely Oort rotation, it follows from formula \ref{eq:beta_oorts} that the main axes of the deformation velocity tensor will be directed at an angle $\beta = 45^\circ$ to the axes of the local rectangular Galactic coordinate system. In this case, the angle $\beta$, for any non-zero values of $A$, will be equal to $45^\circ$. If $C$ or $K$ is non-zero, this means that the rotation is not Oort (axisymmetric). In this case, the angle $\beta \neq 45^\circ$.

\section{Solution of the problem and analysis of the obtained results.}
\label{sec:analysis}

The term vertex is commonly referred to as a point on the sky, relative to which the stellar system under consideration rotates. To determine the coordinates of the vertex, we used the deformation velocity tensors derived by expanding the velocity field of various subsamples (stellar systems) described in Section \ref{sec:vertexcoords}.

\subsection{Using the Gaia DR3 trigonometric parallaxes}
\subsubsection{Determination of the angular vertex coordinates}

As shown above, with axisymmetric (Oort) rotation, the angle $l_{xy}= (\beta-45^\circ)$, called the deviation of the vertex longitudes from the direction to the Galactic center, is exactly equal to zero and does not depend on the coordinate angle $\theta$. The angle $\beta$ in this case is determined by formula \ref{eq:beta_Mp}. 

\begin{figure}
   \centering
\resizebox{\hsize}{!}
   {\includegraphics{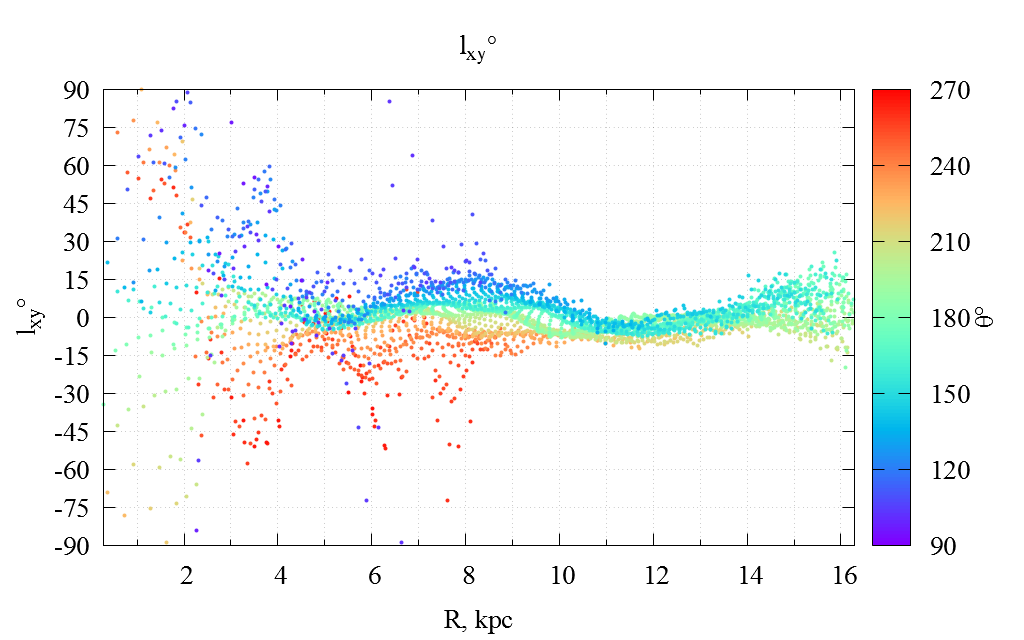}}  
   \caption{Vertex deviations depending on Galactocentric cylindrical coordinates ($R, \theta$).}
\label{fig:lxy_R0=8.28_curve}
\end{figure}

\begin{figure}
   \centering
\resizebox{\hsize}{!}
   {\includegraphics{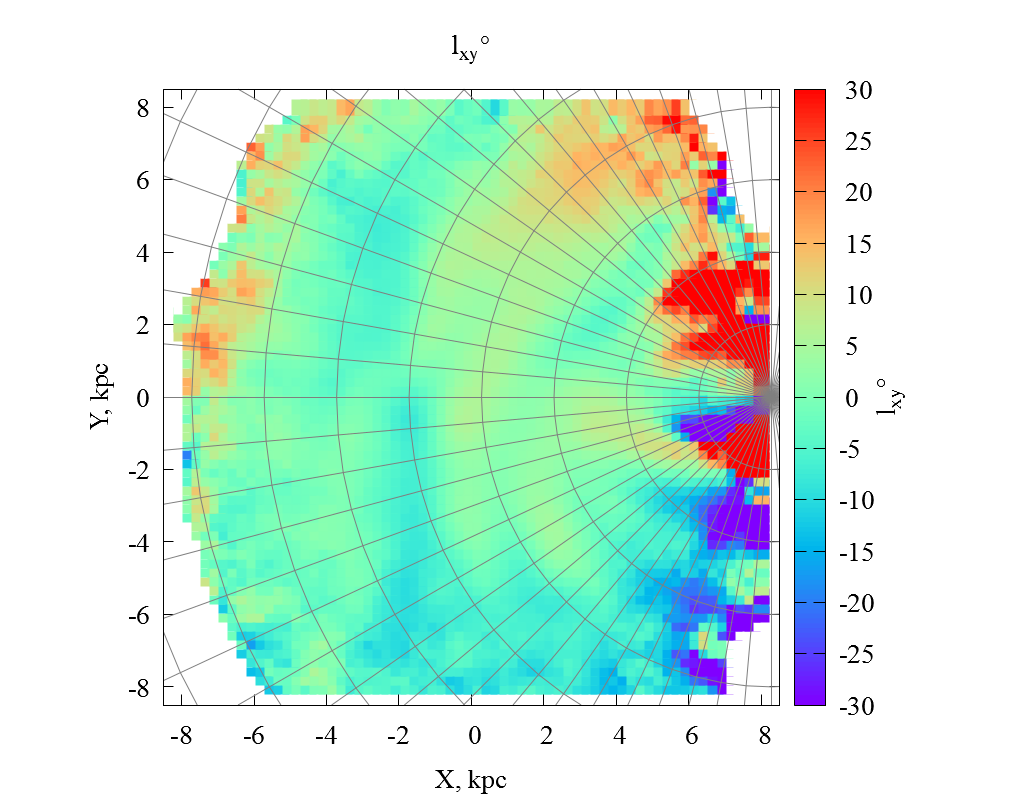}}  
   \caption{Vertex deviations depending on rectangular Galactic coordinates ($X, Y$).}
\label{fig:lxy_R0=8.28_map}
\end{figure}

Fig. \ref{fig:lxy_R0=8.28_curve} shows the dependencies of the vertex deviations $l_{xy}$ on the cylindrical coordinates $R, \theta$. As can be seen from the figure, the values of the angles $l_{xy}$ depending on $R$ for various $\theta$ are noticeably different. While in Fig. \ref{fig:lxy_R0=8.28_map} we show the same values, but in the form of a map, where the rectangular Galactic coordinates $X$ and $Y$ are plotted along the axes, and the $l_{xy}$ value is displayed in color. The advantage of such a graphical presentation of the results is that one can immediately see the behavior of $l_{xy}$ in the entire range of rectangular Galactic coordinates $X$ and $Y$, where the kinematic analysis has been carried out.

In Fig.~\ref{fig:lxy_R0=8.28_curve} it is clearly seen that the angle $l_{xy}$ is not equal to zero, and it changes with $R$ and $\theta$. One can clearly see the “stratification and intertwining” of the dependencies $l_{xy}(R)$ corresponding to various fixed values of the angle $\theta$. This behavior of the dependencies $l_{xy}(R,\theta)$ is probably due to the difference in the deformation velocities in different parts of the Galaxy. This assumption is confirmed by Fig.~\ref{fig:lxy_R0=8.28_map}, which demonstrates the differences in the orientations of the tensor surfaces due to the difference in the deformation velocities in different parts of the Galaxy. However, the similarity of the dependencies $l_{xy}(R)$ in Figs. \ref{fig:lxy_R0=8.28_curve}, at different values of the angle $\theta$, as well as a noticeable predominance of red color in the upper part of Fig.~\ref{fig:lxy_R0=8.28_map}, and blue at the bottom, suggests that "stratification and intertwining" may not be due to kinematic causes alone.

One of these reasons may be an incorrect value of the accepted Galactocentric distance of the Sun $R_\odot$ = 8.28~kpc. In this case, the use of the accepted value of $R_\odot$ to determine the rotation angles of the $x$ axes of local coordinate systems in the direction of the Galactic center will cause an inaccuracy in determining these angles. As a result, the behavior of $l_{xy}(R,\theta)$ will be determined not only by kinematic differences, but also by the orientation inaccuracy of the axes of local coordinate systems.

Checking this assumption, we found that the maximum convergence of the functions $l_{xy}(R,\theta)$ is realized at a certain value $R_{\rm V}$, which differs noticeably from the accepted one. In fact, when using the vertex coordinates to set the orientation of local coordinate systems, the dependencies $l_{xy}(R,\theta)$ become closest and practically turn into one, “least stratified” function $l_{xy}(R)$, which weakly depends on $\theta$.

It also turned out that the use of one vertex does not provide the best convergence of the functions $l_{xy}(R,\theta)$ in the entire range of distances $R$ used. Thus, achieving the best convergence of the functions $l_{xy}(R,\theta)$ in the range of 5--10 kpc results in a noticeable deterioration in convergence within the range of 10--15 kpc (see Fig.~\ref{fig:lxy_usedG_curves} below). This result means that the vertices of different star systems are at different distances from the Sun.

\cite{Amendt1991, Kuijken1991, Smith2012} showed that in a stationary, axisymmetric disk galaxy, the axes of the stellar velocity ellipsoid of any local stellar system ideally coincide with the galactic coordinate axes (e.g., \cite{Binney2008, Smith2012}). The main axes of the deformation velocity tensor for the Oort (axisymmetric) rotation, as shown above, rotate relative to the local axes by an angle of 45$^\circ$. It was shown by \cite{Dehnen2000, Minchev2010, Vorobyov2008, Saha2013} that structures which are not axisymmetric can have a noticeable effect on the observed orientation of the stellar velocity ellipsoids. They will have a similar effect on the orientation of the principal axes of the deformation velocity tensor. 

Therefore, in the general case, when a non-axisymmetric rotation is realized, the direction to the Galactic center (a point on the celestial sphere with coordinates $\alpha_{GC}= 266^\circ$,40499, $\delta_{GC}=-28^\circ$,93617, accepted by the Hipparcos consortium \cite{Perryman1997}) and the direction to the vertex (the point of the celestial sphere relative to which the stellar system rotates) do not coincide. Our result shows that not only the spherical coordinates of the Galactic center and the vertices of stellar systems, but also their distances from the Sun do not coincide.

\subsubsection{Determination of distances to vertices}

The values of the $R_{\rm V}$ distance can be estimated using the approach we proposed. Assuming that the accepted value $R_\odot$=8.28~kpc is correct, the $x$ axes of local coordinate systems will always be directed to the same point -- the Galactic center with coordinates $L=0^\circ$, $B=0^\circ$, $R_\odot$=8.28~kpc. In this case, the main axis $x'$ of the deformation velocity tensor, which is calculated in the local coordinate system $XOY$, being rotated by 45 degrees, will be directed to the vertex. In other words, the longitude of the vertex in the local coordinate system will be numerically equal to the angle $l_{xy}$. Using the values of the angles $l_{xy}$ calculated for local systems, it is possible to construct rays that pass through their vertex point, with their origins locating in the centroids. 

To set the equations of rays (straight lines) passing through two points, we use rectangular Galactic coordinates of specific centroids and points lying on unit circles built around these centroids in the $XOY$ plane. If their radii $r$ are taken equal to 1 kpc, then the coordinates of the second point can be found as follows: $X=X_{c}+r\,{\rm cos}\,l_{xy}$, $Y=Y_{c}+r\,{\rm sin}\,l_{xy}$. Finding the coordinates of the intersection point of two arbitrary rays (straight lines) allows one to find the distance from the Sun to this point in the $XOY$ coordinate system. Pairwise intersections of rays form a certain region of intersection in the Galactic plane.
\begin{figure}
   \centering
\resizebox{\hsize}{!}
   {\includegraphics{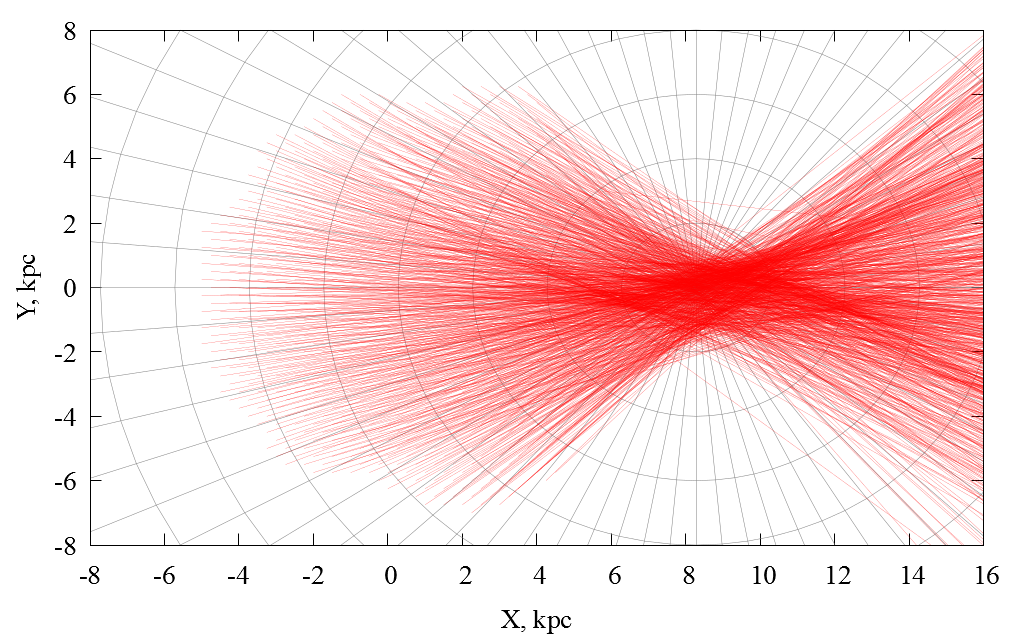}}
\resizebox{\hsize}{!}
   {\includegraphics{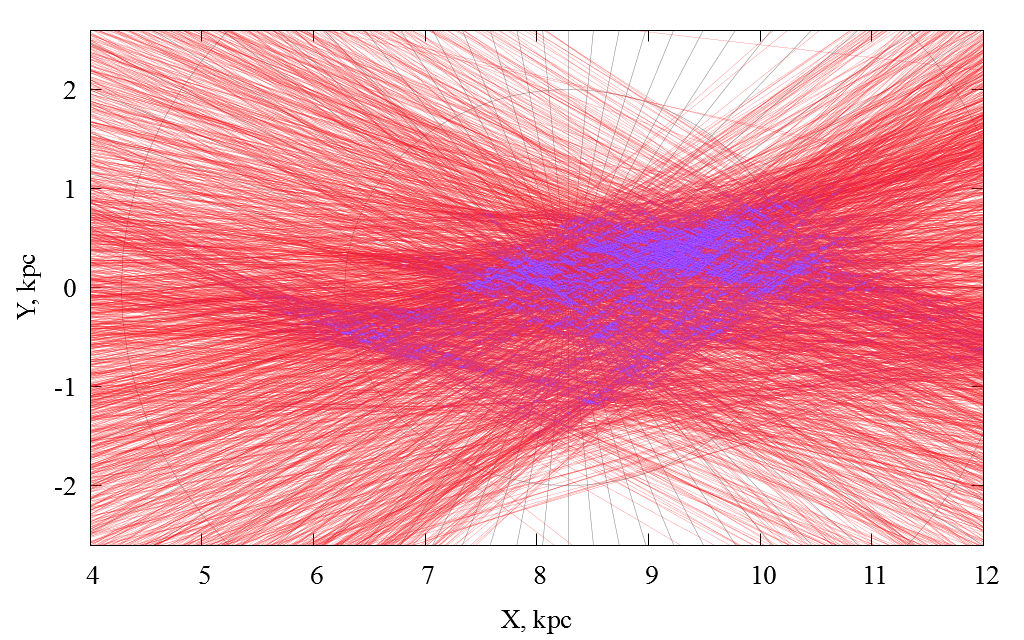}} 
   \caption{Rays directed to the vertex in the Galactic plane and the area of their intersection.}
\label{fig:vertexRays}
\end{figure}
Fig.~\ref{fig:vertexRays} show the region of intersection of the rays. Its rather large sizes and non-uniformity are visible (there are many peaks or nodes). 
The largest of these ray intersection nodes is located approximately at a distance of 9.3~kpc from the Sun. At the same time, it is not located on the $OX$ axis of the rectangular Galactic coordinate system, but it is 0.5~kpc away from it along the $OY$ axis in the positive direction. Similarly, we can select other nodes that can be seen in Fig.~\ref{fig:vertexRays}. The presence of many nodes can be explained by the present nonlinear dependence $l_{xy}(R,\theta)$, which was considered earlier.

The coordinates of the vertices $G^j_{\rm V}(X^j_{\rm V},Y^j_{\rm V})$ , at which the functions $ l_{xy}(R, \theta)$ will have the best convergence in certain ranges of Galactocentric distances $R$, were estimated using the least squares method. To this end, a system of equations was compiled for those rays whose origins got inside the chosen range of Galactocentric distances $\Delta R$. An additional condition for including the ray equation into the system of equations was the presence of at least 25 thousand stars in the stellar system. The solution of the system was the desired coordinates of the point of intersection of all rays --- the vertex. 

So, for the range $7<R<10 kpc$, we got point $G^1_{\rm V}$ with coordinates: $X^1_{\rm V}=9.82$~kpc, $Y^1_{\rm V}=0.31$~kpc. For the range of Galactocentric distances $10 < R < 15$~kpc, we got another point -- $G^2_{\rm V}$, which already has different coordinates: $X^2_{\rm V}=8.99$~kpc, $Y^2_{\rm V}=-0.36$~kpc. Also, we have estimated the coordinates of the point $G^0_{\rm V}=G_{\rm V}$, obtained using the entire available range of $R$ and called by us the general vertex. The coordinates of this point are:  $X^0_{\rm V}=9.39$~kpc, $Y^0_{\rm V}=0.59$~kpc.

In table \ref{tab:vertices}, we provide for all points $G^j_{\rm V}$ the coordinates and errors of their determination, as well as the calculated heliocentric distances.
\begin{table}\centering
\caption{Results of estimation of Galactic rectangular coordinates of vertices}
\label{tab:vertices}\begin{tabular}{l|c|c|c}
\hline
 & $G^0_{\rm V}=G_{\rm V}$ & $G^1_{\rm V}$ & $G^2_{\rm V}$ \\
\hline
$\Delta R$, kpc      & 0--16 & 7--10 & 10--15 \\
$X_V$, kpc           & 9.39  & 9.82  & 8.99  \\
$Y_V$, kpc           & 0.59  & 0.31  & -0.36 \\
$R_V$, kpc           & 9.39  & 9.83  & 8.99  \\
$\epsilon(X_V)$, kpc & 0.04  & 0.04  & 0.07  \\
$\epsilon(Y_V)$, kpc & 0.02  & 0.02  & 0.03  \\
\hline
\end{tabular}
\end{table}

\begin{figure}
   \centering
\resizebox{\hsize}{!}
   {\includegraphics{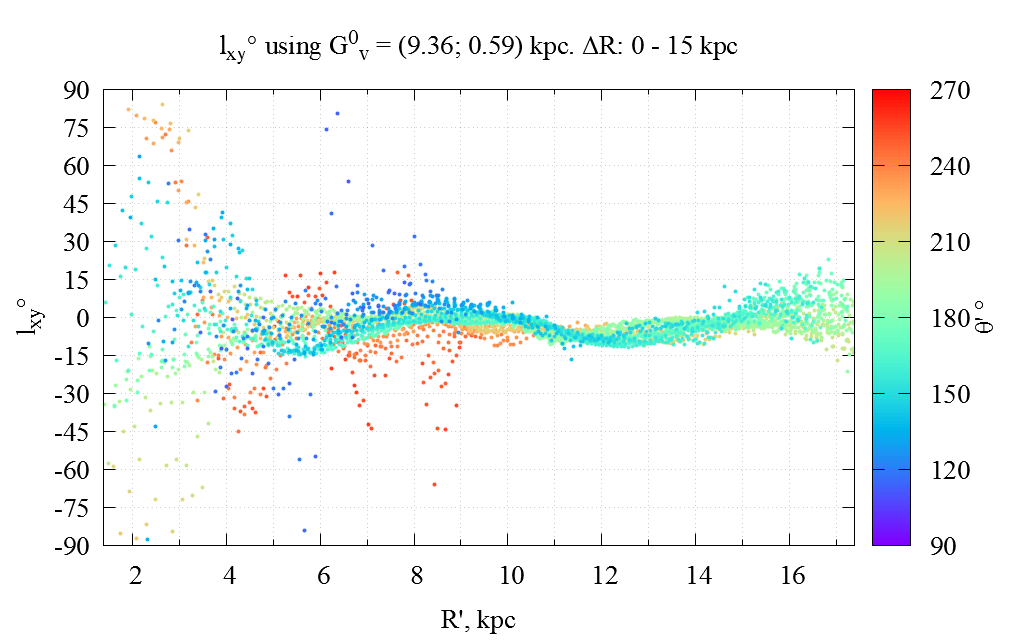}}
\resizebox{\hsize}{!}
   {\includegraphics{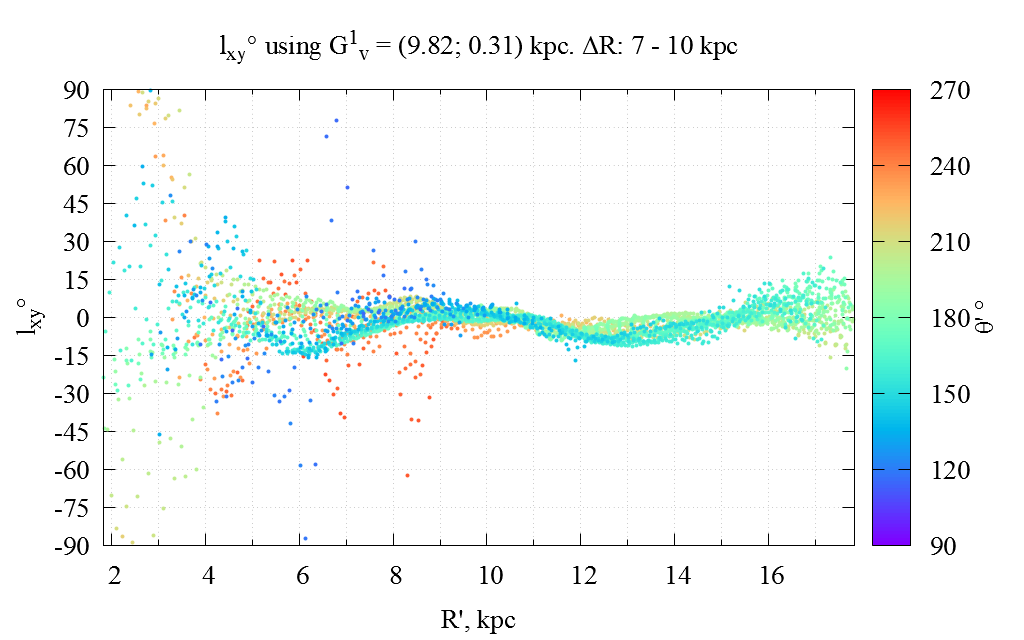}} 
\resizebox{\hsize}{!}
   {\includegraphics{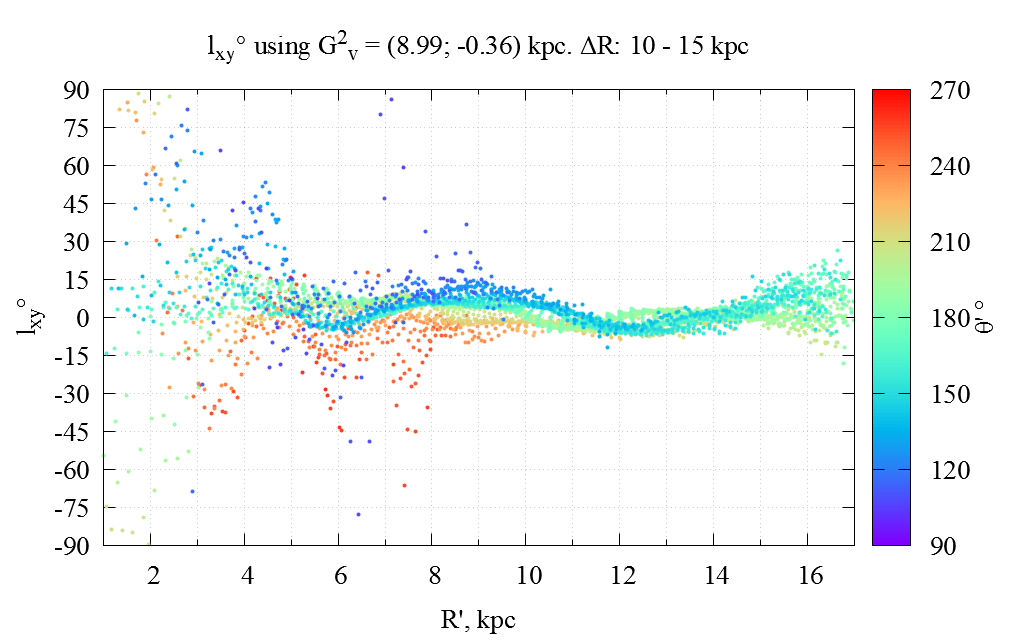}}       
   \caption{Vertex deviation depending on “vertexcentric” cylindrical coordinates. The determined vertex coordinates have been used to set the orientation of local Galactic rectangular coordinate systems.}
\label{fig:lxy_usedG_curves}
\end{figure}

\begin{figure}
   \centering
\resizebox{\hsize}{!}
   {\includegraphics{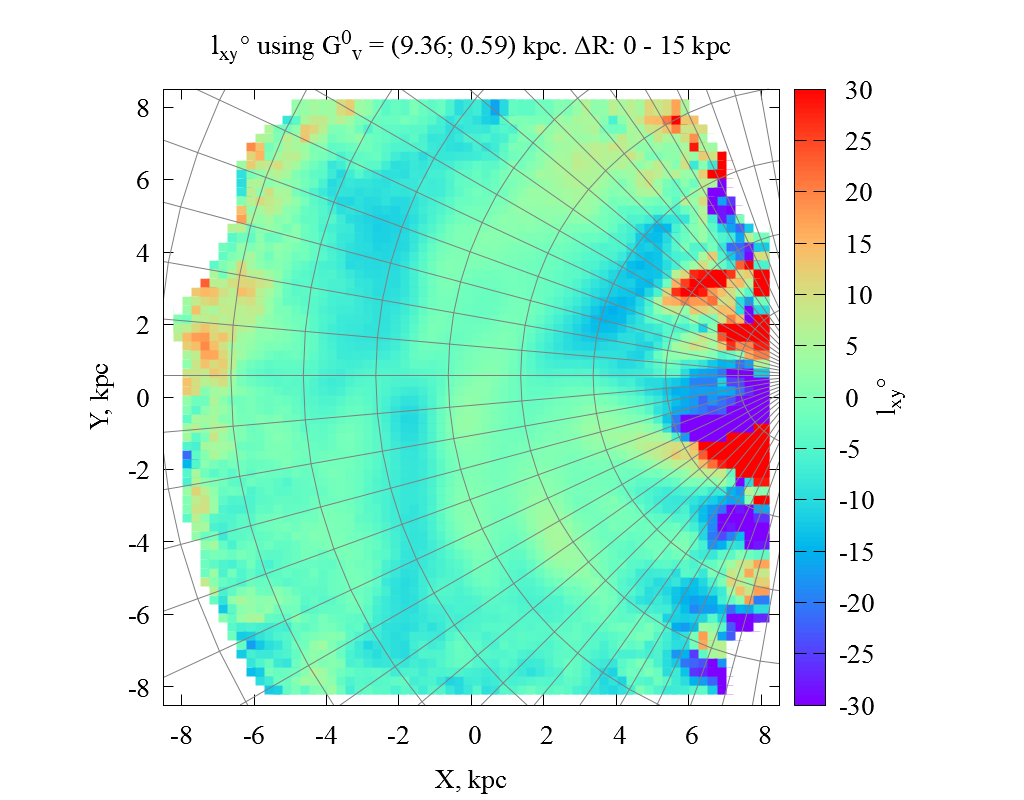}}
\resizebox{\hsize}{!}
   {\includegraphics{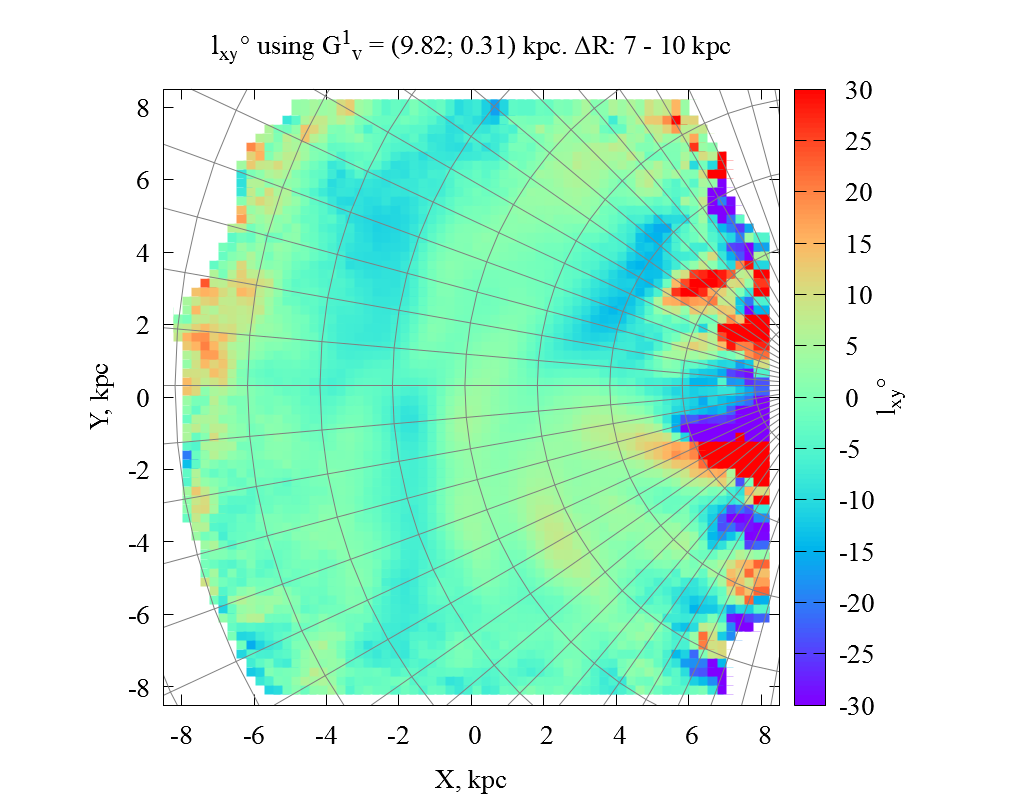}} 
\resizebox{\hsize}{!}
   {\includegraphics{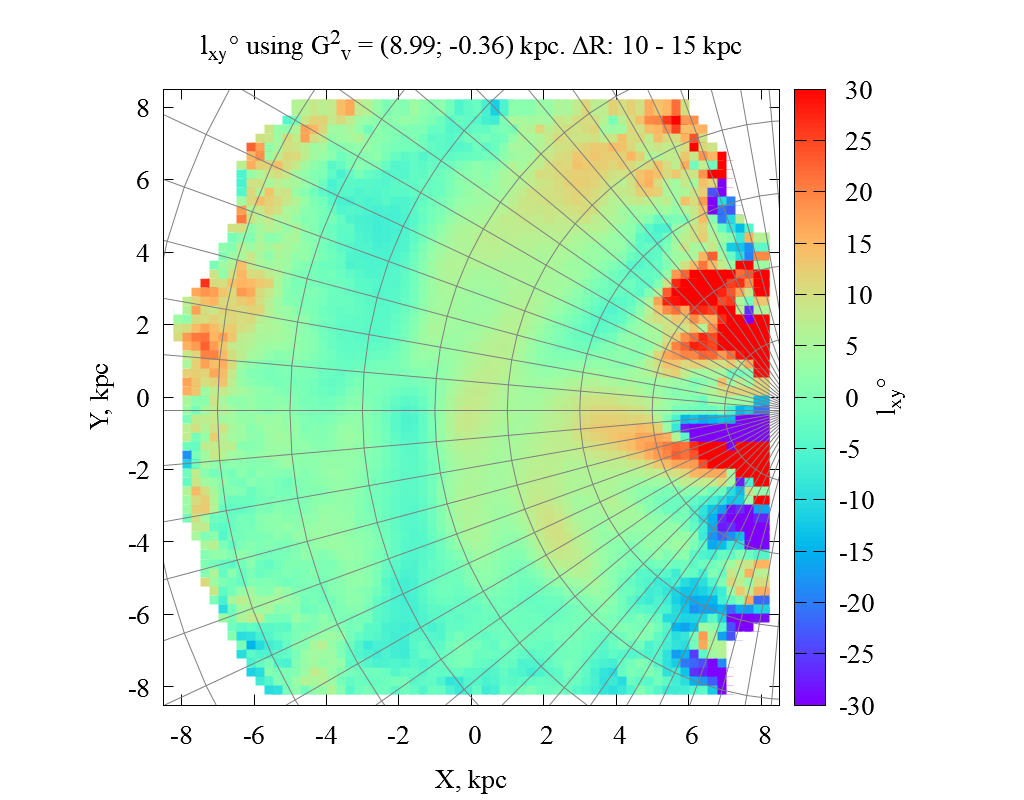}}  
   \caption{Vertex deviation depending on rectangular Galactic coordinates.The determined vertex coordinates have been used to set the orientation of local Galactic rectangular coordinate systems.}
\label{fig:lxy_usedG_maps}
\end{figure}

Now we can use the coordinates of the point $G^j_{\rm V}(X, Y)$ instead of the $X$-th coordinate of the Galactic center, equal to $R_\odot$. This allows us to define new “vertexcentric” cylindrical coordinate systems $R', \theta'$ and orient properly the local coordinate system in order to construct new dependencies $l_{xy}(R',\theta')$ and $l_{xy}(X,Y)$. These results are shown in Figs. \ref{fig:lxy_usedG_curves} and \ref{fig:lxy_usedG_maps}.

Comparing Figs. \ref{fig:lxy_R0=8.28_curve} and \ref{fig:lxy_usedG_curves}, one can note an improvement in the convergence of the dependence $l_{xy}(R',\theta')$ in those ranges of $R$ where the corresponding $G^j_{\rm V}$ was used. A similar conclusion can be reached by comparing Figs. \ref{fig:lxy_R0=8.28_map} and \ref{fig:lxy_usedG_maps}. It should be noted that the local structures, which are clearly visible on the maps, remained the same, which indirectly confirms their purely kinematic, and not geometric, nature. The deviation of the angle $l_{xy}$ in the entire region under study from zero can be considered as a measure of the non-axisymmetry of the Galaxy.

\subsection{Using corrected parallaxes and proper motions of Gaia DR3}

It is known from \cite{Lindegren2021} the $Gaia$ DR3 parallaxes are biased, that is, they are systematically offset from the expected distribution around zero by several tens of microarcseconds. The authors show that this parallax offset depends in a non-trivial way on magnitude, color, and ecliptic latitude of the source, and give recommendations for a systematic correction of the parallaxes. In particular, users are strongly encouraged to make their own judgment as to whether the specified offset correction is appropriate for their particular application.

To correct parallaxes, we used the Parallax bias $Z_5$ computed according to the table 9 in the paper by \cite{Lindegren2021}. The distribution of the number of stars in our sample is limited to about magnitude 17 due to they were selected taking into account the presence of radial velocity measurements. Therefore, we divided the range of stellar magnitudes into only two parts. In the magnitude range from 6 to 13, we applied an offset of -30 $\mu$as, and -40 $\mu$as in the range $m >$ 13. Since in the work by \cite{Lindegren2021} in Fig. 20, jumps are seen in the range from 11.5 to 13 magnitudes, we decided to use an average value of approximately $Z_5$ = -30 $\mu$as in this range.

In addition, we have corrected the proper motions of stars in the magnitude range from 6 to 13. It was already shown by \cite{Fedorov2018} that the proper motions of TGAS stars in the range from 11.5 to 13 magnitudes were distorted by the magnitude equation. \cite{Lindegren2018, Brandt2018} showed that in the second release of $Gaia$ (DR2) data, the reference frame of bright stars rotates at a velocity of $\sim$0.15 \myr\, relative to faint stars and quasars. In EDR3, this rotation was previously removed (see Section 4.5 in \cite{Lindegren2021}).

To align the proper motions system of stars in our sample brighter than $G=13$ with the International Celestial Coordinate System, we used corrections calculated according to the recommendations presented by \cite{Cantat-Gaudin2021}. 

\begin{figure}
   \centering
\resizebox{\hsize}{!}
   {\includegraphics{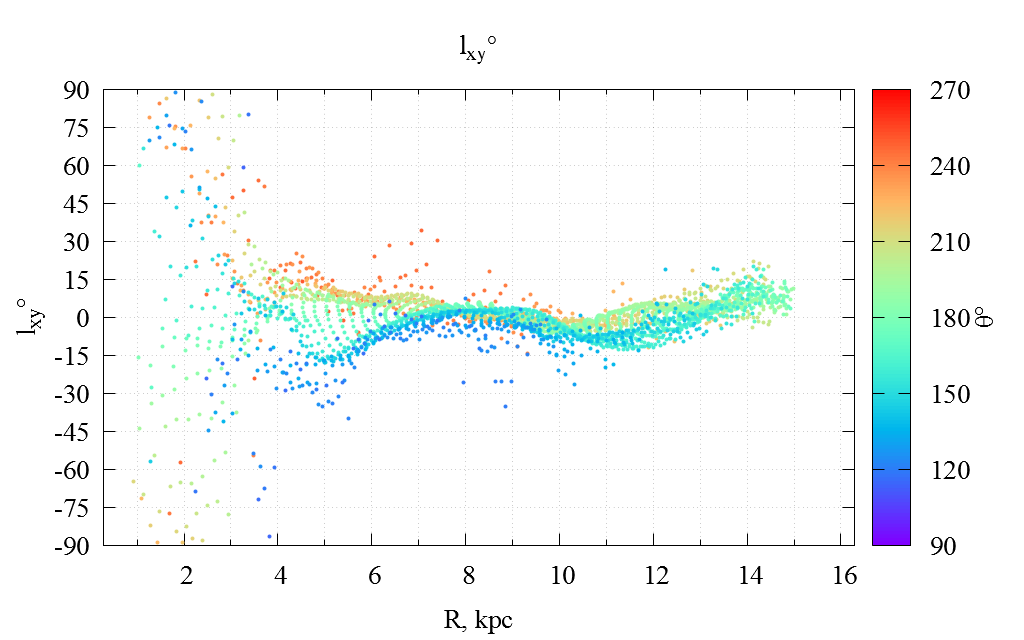}}  
   \caption{Vertex deviations depending on Galactocentric cylindrical coordinates ($R, \theta$). Corrected parallaxes (\protect\cite{Lindegren2021}) and proper motions (\protect\cite{Cantat-Gaudin2021}) have been used.}
\label{fig:lxy_R0=8.28_curve_rLINmuTr}
\end{figure}

\begin{figure}
   \centering
\resizebox{\hsize}{!}
   {\includegraphics{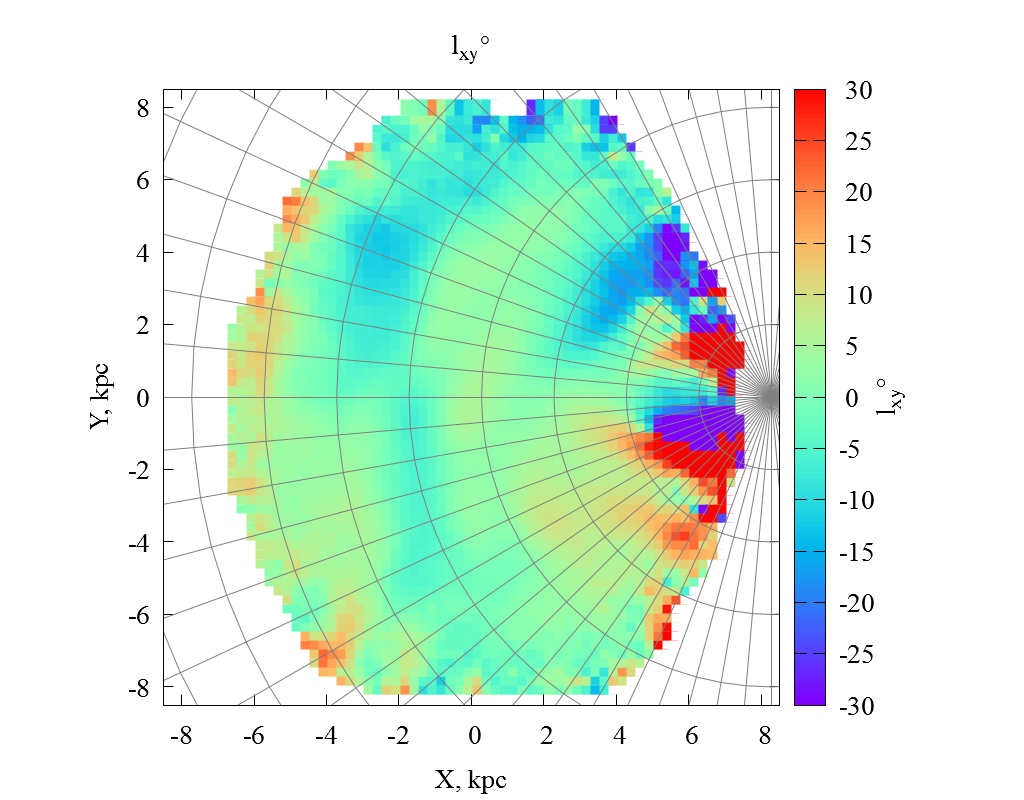}}  
   \caption{Vertex deviations depending on rectangular Galactic coordinates ($X, Y$). Corrected parallaxes (\protect\cite{Lindegren2021}) and proper motions (\protect\cite{Cantat-Gaudin2021}) have been used.}
\label{fig:lxy_R0=8.28_map_rLINmuTr}
\end{figure}

\begin{figure}
   \centering
\resizebox{\hsize}{!}
   {\includegraphics{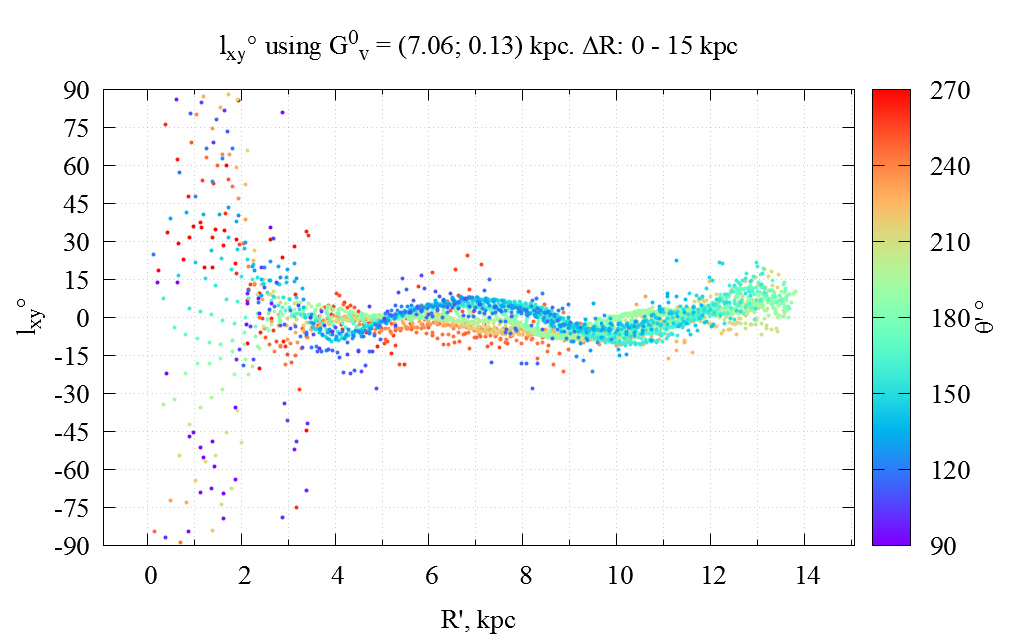}}
\resizebox{\hsize}{!}
   {\includegraphics{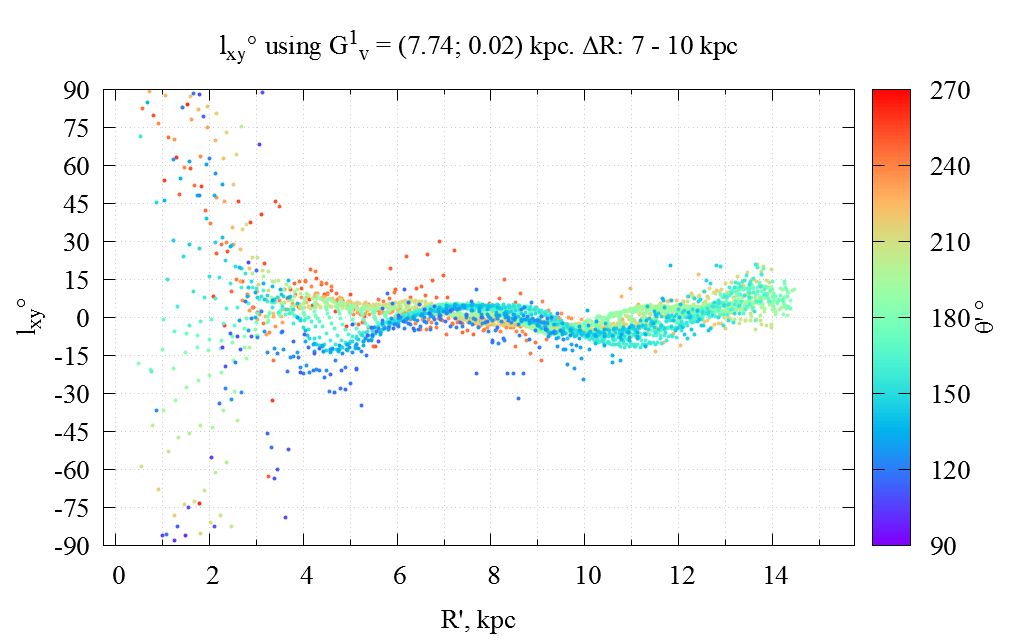}} 
\resizebox{\hsize}{!}
   {\includegraphics{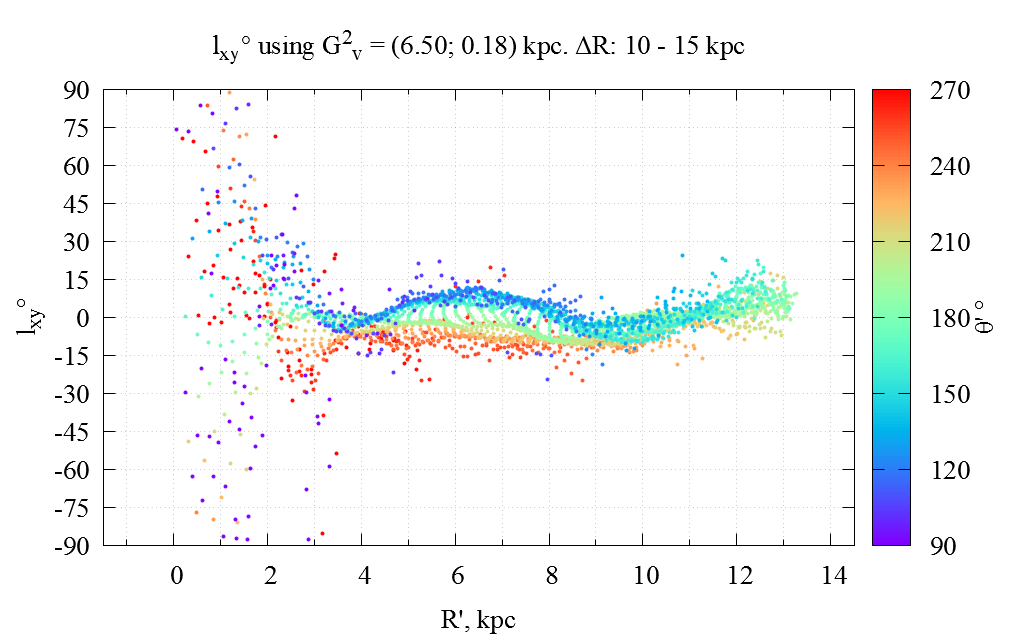}}       
   \caption{Vertex deviation depending on “vertexcentric” cylindrical coordinates. The determined vertex coordinates have been used to set the orientation of local Galactic rectangular coordinate systems. Corrected parallaxes (\protect\cite{Lindegren2021}) and proper motions (\protect\cite{Cantat-Gaudin2021}) have been used.}
\label{fig:lxy_usedG_curves_rLINmuTr}
\end{figure}

\begin{figure}
   \centering
\resizebox{\hsize}{!}
   {\includegraphics{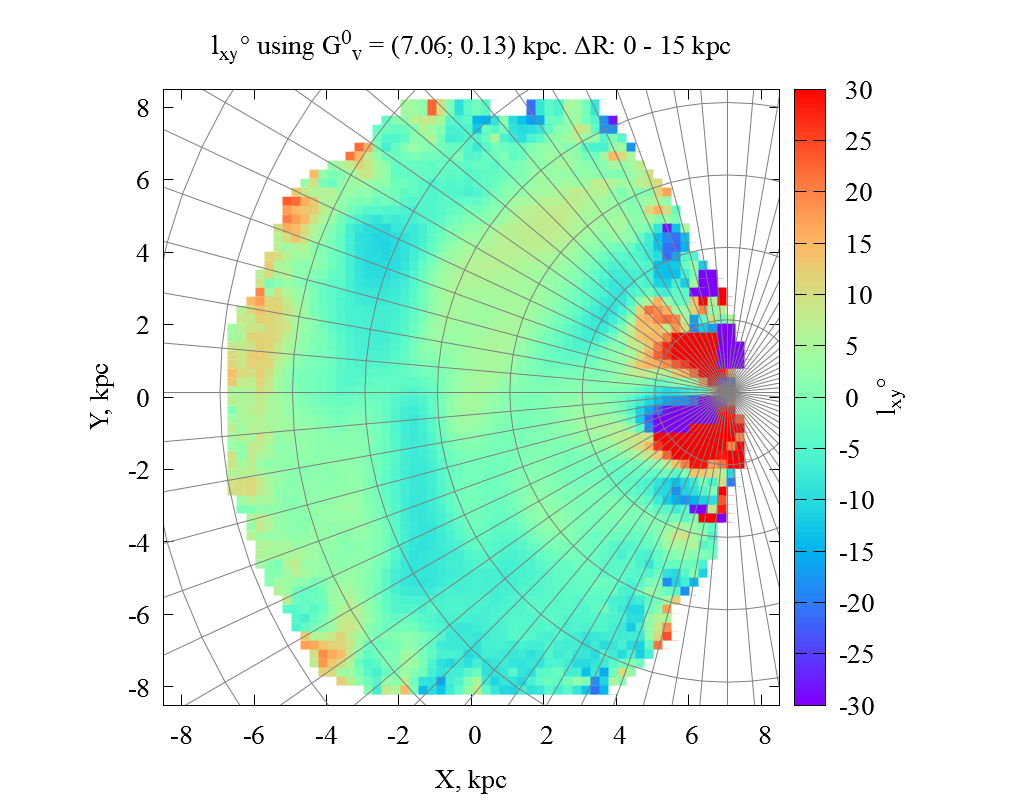}}
\resizebox{\hsize}{!}
   {\includegraphics{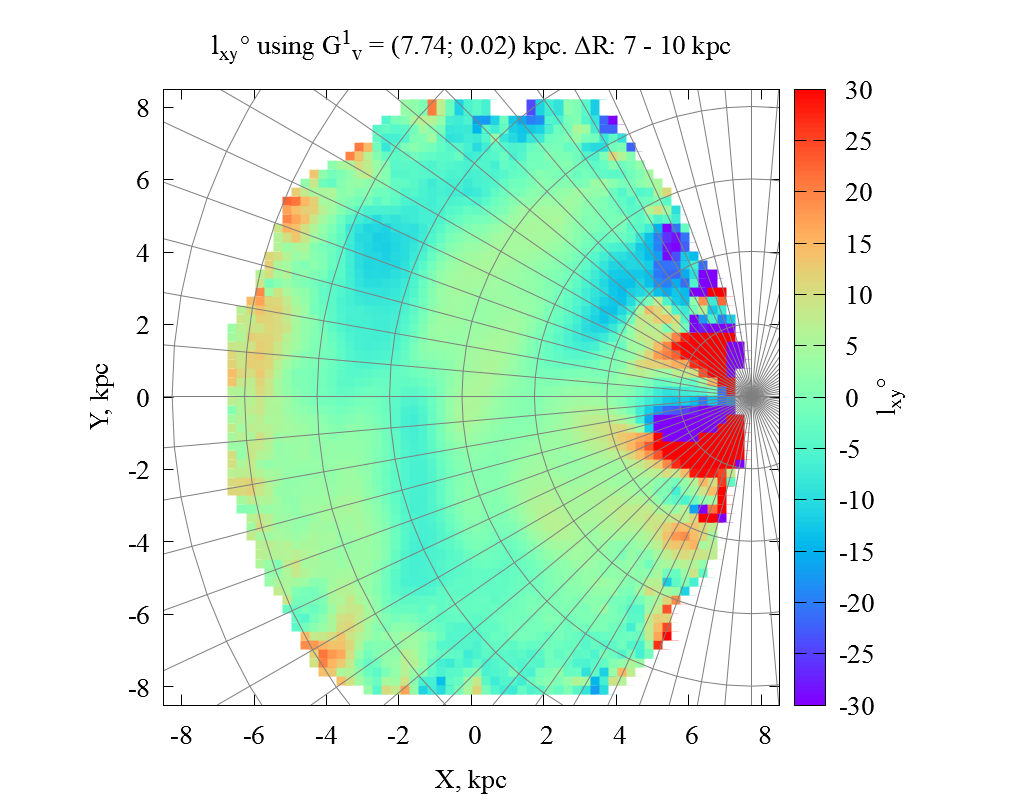}} 
\resizebox{\hsize}{!}
   {\includegraphics{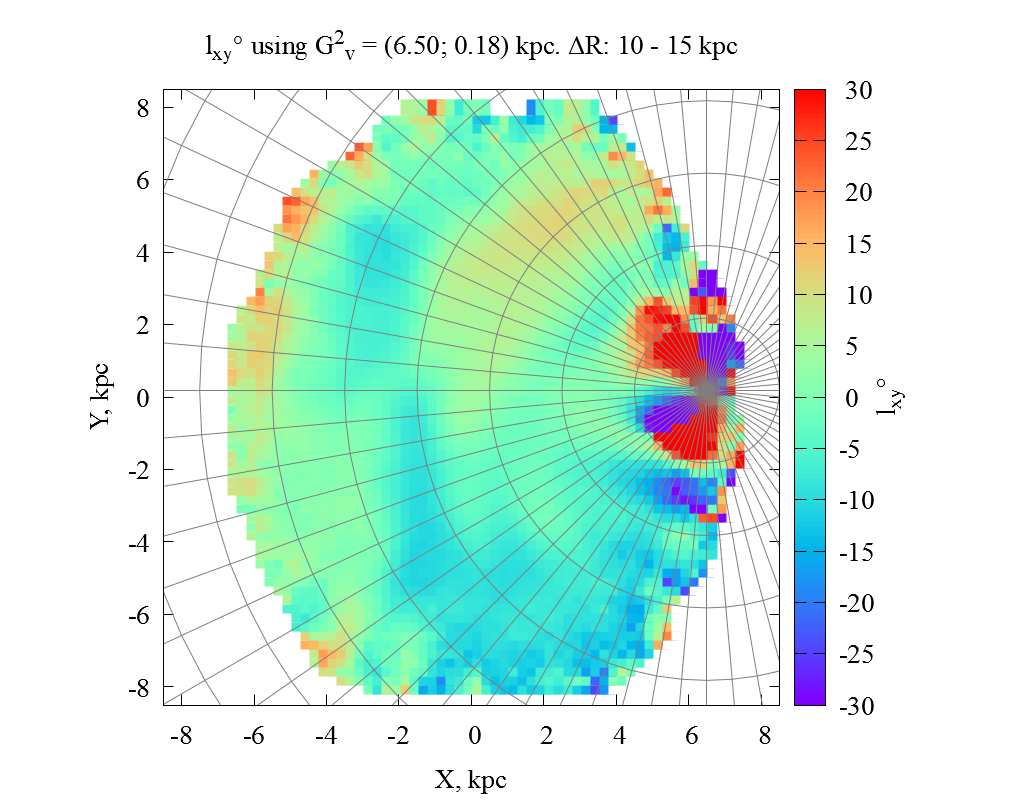}}   
   \caption{Vertex deviation depending on rectangular Galactic coordinates.The determined vertex coordinates have been used to set the orientation of local Galactic rectangular coordinate systems. Corrected parallaxes (\protect\cite{Lindegren2021}) and proper motions (\protect\cite{Cantat-Gaudin2021}) have been used.}
\label{fig:lxy_usedG_maps_rLINmuTr}
\end{figure}

\begin{figure}
   \centering
\resizebox{\hsize}{!}
   {\includegraphics{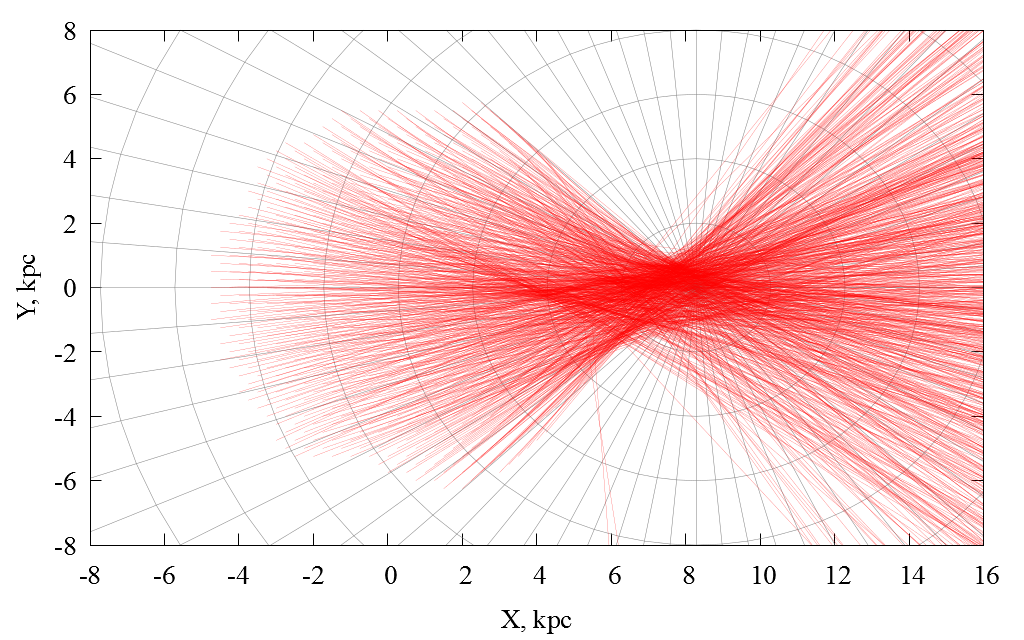}}
\resizebox{\hsize}{!}
   {\includegraphics{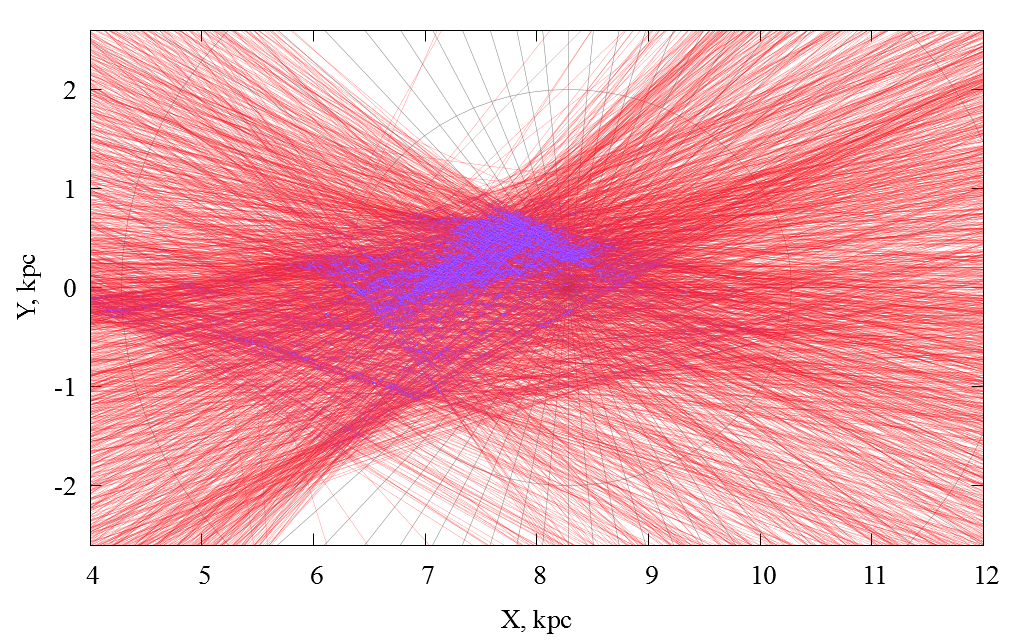}} 
      \caption{Rays directed to the vertex in the Galactic plane and the area of their intersection, using corrected parallaxes (\protect\cite{Lindegren2021}) and proper motions (\protect\cite{Cantat-Gaudin2021}).}
\label{fig:vertexRays_rLINmuTr}
\end{figure}

The figures \ref{fig:lxy_R0=8.28_curve_rLINmuTr}, \ref{fig:lxy_R0=8.28_map_rLINmuTr}, \ref{fig:lxy_usedG_curves_rLINmuTr}, \ref{fig:lxy_usedG_maps_rLINmuTr}, \ref{fig:vertexRays_rLINmuTr} and tables \ref{tab:vertices_rLINmuTr}, \ref{tab:vertices_rPhotogeo} show the results obtained after applying the corrections of parallaxes and proper motions to our stellar sample. We provide plots with dependencies similar to those presented in 4.1.1 and 4.1.2, however, we omit the description of some figures, since the general idea of the behavior of the values given on them remains unchanged. At the same time, changes in these quantities eventually lead to some change in the dependence $l_{xy}(R',\theta')$ . This, in turn, leads to the fact that the best convergence of the functions $l_{xy}(R',\theta')$ is realized for some other $G_V$ values, which are noticeably smaller than the $G_V$ found in the previous subsection. The results of determining the vertex coordinates in this case are presented in Table \ref{tab:vertices_rLINmuTr}.


\begin{table}\centering
    \caption{Results of estimation of Galactic rectangular coordinates of vertices, using corrected parallaxes (\protect\cite{Lindegren2021}) and proper motions (\protect\cite{Cantat-Gaudin2021}).}
\label{tab:vertices_rLINmuTr}\begin{tabular}{l|c|c|c}
\hline
 & $G^0_{\rm V}=G_{\rm V}$ & $G^1_{\rm V}$ & $G^2_{\rm V}$ \\
\hline
$\Delta R$, kpc      & 0--16 & 7--10 & 10--15 \\
$X_V$, kpc           & 7.06  & 7.74  & 6.5  \\
$Y_V$, kpc           & 0.13  & 0.02  & 0.18 \\
$R_V$, kpc           & 7.07  & 7.74  & 6.51  \\
$\epsilon(X_V)$, kpc & 0.03  & 0.04  & 0.08  \\
$\epsilon(Y_V)$, kpc & 0.02  & 0.03  & 0.04  \\
\hline
\end{tabular}
\end{table}

At the same time, the distances to the vertices have changed by about 2 kpc. This result is not unexpected and does not change our conclusion that both the angular coordinates of the Galactic center and the vertices of stellar systems, as well as their distances from the Sun do not coincide.

For comparison, in Table \ref{tab:vertices_rPhotogeo} we also present the results obtained after using the photogeometric distances from \cite{Bailer-Jones2021} and correcting the stellar proper motion in our sample for the magnitude equation. In order not to overload the paper with illustrative materials, we do not provide the corresponding figures, but only note that the behavior and values of the parameters given on them are in good agreement with those presented in the figures \ref{fig:lxy_R0=8.28_curve_rLINmuTr}-\ref{fig:vertexRays_rLINmuTr}.

\begin{table}\centering
\caption{Results of estimation of Galactic rectangular coordinates of vertices, using photogeometric distances to stars (\protect\cite{Bailer-Jones2021}) and corrected proper motions (\protect\cite{Cantat-Gaudin2021})}
\label{tab:vertices_rPhotogeo}\begin{tabular}{l|c|c|c}
\hline
 & $G^0_{\rm V}=G_{\rm V}$ & $G^1_{\rm V}$ & $G^2_{\rm V}$ \\
\hline
$\Delta R$, kpc      & 0--16 & 7--10 & 10--15 \\
$X_V$, kpc           & 7.06  & 8.07  & 6.6  \\
$Y_V$, kpc           & -0.10 & 0.15  & 0.15  \\
$R_V$, kpc           & 7.07  & 8.07  & 6.6  \\
$\epsilon(X_V)$, kpc & 0.03  & 0.04  & 0.07  \\
$\epsilon(Y_V)$, kpc & 0.02  & 0.02  & 0.04  \\
\hline
\end{tabular}
\end{table}

\section{Summary and Conclusions}
\label{sec:conclusions}

The use of $Gaia$~DR3 data makes it possible to determine the vertex coordinates of various stellar regions whose centroids are located in the Galactic plane at heliocentric distances up to 8–-10 kpc. The approach proposed in the works by \cite{Fedorov2021, Fedorov2023} for the analysis of kinematic parameters, when a fictitious observer, being at an arbitrary point in the Galaxy, determines them in the framework of the O--M model, made it possible to obtain a number of new results related to a significant part of the Galaxy. 

In this work, we have applied this approach to determine the coordinates of the vertices of various stellar systems contained in spherical regions with a radius of 1 kpc, whose centers are located in the Galactic plane. Since the analysis revealed that the $M^+$ deformation velocity tensors calculated in local coordinate systems are almost flat, we used only four components $m^+_{11}$, $m^+_{12}$, $m^+_{21}$, $m^+_{22}$. It turned out that the deviations of the vertices of these stellar systems obey a certain law and are presented mainly in the form of the dependence $l(R)$ and weakly depend on $\theta$. This result could not have been obtained without knowledge of the spatial coordinates and velocities of the stars contained in $Gaia$ DR3 and allowing one to set the local Galactic coordinate system at an arbitrary point in the Galactic plane.

Usually, in works on determining the deviations of the vertex, it is explicitly or implicitly assumed that the distance from the Sun to the Galactic center and to the vertex are the same. This is indeed true for axisymmetric systems. For the case when $m^+_{11}$ and $m^+_{22}$ are not equal to zero, we show that the vertices of different stellar systems are located at different distances that do not coincide with the accepted distance of the Sun $R_\odot$=8.28 kpc to the center of the Galaxy. This indicates that for the investigated part of the Galaxy there is no single center of rotation, as in the case of axisymmetric systems. 
Unfortunately, our approach does not allow us to indicate the exact heliocentric distances of the vertices, since they turn out to be dependent on the knowledge of the systematic errors of the parallaxes used. Although we cannot give exact values of the distances from the Sun to the vertices, it is still possible to indicate suitable values of $G_{\rm V}$, at which the function $l_{xy}(R',\theta')$, in a specific range of Galactocentric distances, has a minimum stratification.

Our results indicate that not only the angular coordinates of the Galactic center and the vertices of stellar systems do not coincide, but also their distances to the Sun do not coincide either. These results can be useful in many kinematic and dynamic problems.

\section{Acknowledgements}
This work has made use of data from the European Space Agency (ESA) mission {\it Gaia} (\url{https://www.cosmos.esa.int/gaia}), processed by the {\it Gaia} Data Processing and Analysis Consortium (DPAC,\url{https://www.cosmos.esa.int/web/gaia/dpac/consortium}). Funding for the DPAC has been provided by national institutions, in particular the institutions participating in the {\it Gaia} Multilateral Agreement.

We are immensely grateful to the Armed Forces of Ukraine for the fact that in wartime we still have the opportunity to work and do science.

We sincerely thank the anonymous reviewer for a careful reading the paper, very useful comments, and most importantly, for constructive suggestions.

\section*{Data availability}
\addcontentsline{toc}{section}{Data availability}
The used catalogue data is available in a standardised format for readers via the CDS (https://cds.u-strasbg.fr).
The software code used in this paper can be made available upon request by emailing the corresponding author. 


\bsp 
\label{lastpage}
\end{document}